\documentclass[12pt]{article}

\usepackage{amsmath,amssymb,rotating,wasysym}

\usepackage{lscape,chngpage}
 
\def\Journal#1#2#3#4{{#1} {\bf #2}, #3 (#4)}

\def\CQG{\em Class. Quantum Grav.}

\def\PRD{\em Phys. Rev. D }

\def\JMP{\em J. Math. Phys.}

\def\PRL{\em Phys. Rev. Lett.}

\newcommand{\bm}[1]{\mbox{\boldmath $#1$}}
\def\espaitemps{({\cal V},g)}
\def\varietat{{\cal V}}

\def\be{\begin{equation}}
\def\ee{\end{equation}}
\def\bea{\begin{eqnarray}}
\def\eea{\end{eqnarray}}
\def\bean{\begin{eqnarray*}}
\def\eean{\end{eqnarray*}}

\def\unwarrow{\nwarrow\joinrel\joinrel\uparrow}
\def\unearrow{\uparrow\joinrel\joinrel\nearrow}
\def\dswarrow{\swarrow\joinrel\joinrel\downarrow}
\def\dsearrow{\downarrow\joinrel\joinrel\searrow}
\def\dotsearrow{\hspace{-5mm}\dot{\hspace{5mm}\searrow}}
\def\dotswarrow{\hspace{5mm}\dot{\hspace{-5mm}\swarrow}}
\def\dotdswarrow{\swarrow\joinrel\joinrel\joinrel\dot\downarrow}
\def\dotdsearrow{\dot\downarrow\joinrel\joinrel\joinrel\searrow}
\def\dsarrow{\swarrow\joinrel\joinrel\downarrow\joinrel\joinrel\searrow}
\def\unarrow{\nwarrow\joinrel\joinrel\uparrow\joinrel\joinrel\nearrow}

\begin{document}

\title{Classification of spacelike surfaces in spacetime}
\author{Jos\'e M. M. Senovilla \\
F\'{\i}sica Te\'orica, Universidad del Pa\'{\i}s Vasco, \\
Apartado 644, 48080 Bilbao, Spain \\ 
josemm.senovilla@ehu.es}
\date{}
\maketitle
\begin{abstract} 
A classification of 2-dimensional surfaces imbedded in spacetime is presented, according to the algebraic properties of their shape tensor. The classification has five levels, and provides among other things a refinement of the concepts of trapped, umbilical and extremal surfaces, which split into several different classes. The classification raises new important questions and opens many possible new lines of research. These, together with some applications and examples, are briefly considered.
\end{abstract} 


\section{Introduction}
The purpose of this paper is to present a complete classification of two-dimensional spacelike surfaces imbedded in a four-dimensional Lorentzian manifold. The classification will be carried out according to the {\em extrinsic} properties of the surface, that is to say, to the algebraic types of its shape tensor. Equivalently, it is an algebraic classification based, at each point, on the properties of two independent second fundamental forms. In particular, I will use two {\em null} second fundamental forms.

The classification has five levels of increasing complexity:
\begin{enumerate}
\item the first level simply uses the algebraic types of each of the two null second fundamental forms at a given point. It has 8 different cases for each of them, plus some degenerate subcases;
\item the second level relies on the combination of all possible algebraic types of the two null second fundamental forms; hence it has 64 classes plus the degeneracies.
\item the third level refines the previous one by taking into account the relative orientations of the two null second fundamental forms at the chosen point. This adds a (continuous) parameter to the previous cases, with some remarkable types for particular values of this parameter.
\item the fourth level depends on the entire surface and is based on how the sign of the traces of the null second fundamental forms change on the surface ---equivalently, on the causal character and orientation of the mean curvature vector over the surface.
\item finally, the fifth level is also global and uses the variations of the first three levels on the surface.
\end{enumerate}
The first three levels are purely local, valid at each point of the surface. The last two levels are global and rely on the type of points ---classified according to first three levels--- which are {\em missing} on a given surface.

I have tried to unify the nomenclature used in the literature, as well as to devise a graphical, easily remembered, symbol for each type of surface. The usefulness and potential applicability of the entire classification is analyzed in section \ref{sec:apps}.

\subsection{Basic concepts and notation}
Let $\espaitemps$ be a $4$-dimensional\footnote{The generalization to higher dimensional space-times is dealt with in subsection \ref{subsec:higher}} Lorentzian manifold with metric tensor $g$ of signature $(-,+,+,+)$. A surface is an imbedded $2$-dimensional manifold $(S,\Phi)$, where $\Phi : S \longrightarrow \varietat$ is an imbedding \cite{O,HE}. For the sake of brevity $S$ will be identified with its image $\Phi (S)$ in $\varietat$. Without loss of generality I will assume that the surface is connected (otherwise, it is enough to restrict everything that follows to each one of its connected components.) The surface is spacelike if the inherited metric (or first fundamental form) $\bm{\gamma}\equiv \Phi^{*}g$ is positive-definite on $S$, which will be assumed in what follows. Then, at any $x\in S$ one has the orthogonal decomposition of the tangent space
$$
T_{x}\varietat =T_{x}S\oplus T_{x}S^{\perp}\, .
$$
Thus, $\forall x\in S,\,\,\, \forall \vec z\in T_{x}\varietat$ we have $\vec z 
=\vec{z}^{T} + \vec{z}^{\perp}$ which are called the {\em tangent} 
and {\em normal} parts of $\vec z$ relative to $S$. In particular, for all smooth vector fields $\vec X,\vec Y$ tangent to $S$ (i.e., for all local sections of $TS$):
$$
\forall \vec X,\vec Y\in \mathfrak{X}(S),\hspace{1cm}
\nabla_{\vec X}\, \vec Y = \overline\nabla_{\vec X}\, \vec Y
- \vec{\bm{K}}(\vec X,\vec Y)
$$
where $\nabla$ is the connection on $\espaitemps$, $\overline\nabla$ is the inherited connection ---which coincides with the Levi-Civita connection of $\bm{\gamma}$: $\overline\nabla\bm{\gamma}=0$---, and
$$
\overline\nabla_{\vec X}\, \vec Y\equiv \left(\nabla_{\vec X}\, \vec Y\right)^{T}, \hspace{1cm}
-\vec{\bm{K}}(\vec X,\vec Y)\equiv \left(\nabla_{\vec X}\, \vec Y\right)^{\perp}
$$ 
are the tangent and perpendicular parts of $\nabla_{\vec X}\, \vec Y$ relative to $S$, respectively.
The basic object to be used in what follows is precisely 
$$
\vec{\bm{K}} : \, \mathfrak{X}(S) \times \mathfrak{X}(S) \longrightarrow \mathfrak{X}(S)^{\perp}\, ,
$$
called the {\em shape tensor} of $S$ in $\varietat$. 
The shape tensor contains the information concerning the ``shape'' 
of $S$ within $\varietat$ along {\em all} directions normal to 
$S$. Given any normal direction $\vec n\in \mathfrak{X}(S)^{\perp}$, the {\em second fundamental form} of $S$ in $\espaitemps$ relative to $\vec n$ is the 2-covariant symmetric tensor field on $S$ defined by means of
$$
\bm{K}_{\vec n}(\vec X,\vec Y)\equiv g\left(\vec n,\vec{\bm{K}}(\vec X,\vec Y)\right),
\hspace{1cm} \forall \vec X,\vec Y\in \mathfrak{X}(S)\, .
$$
The {\em mean curvature vector} $\vec H$ of $S$ in $\espaitemps$ is an averaged version of the shape tensor defined by
$$
\vec H =\mbox{tr}\,  \vec{\bm{K}} , \hspace{5mm} \vec H \in \mathfrak{X}(S)^{\perp}
$$
where the trace tr is taken with respect to $\bm{\gamma}$. Each 
component of $\vec H$ along a particular normal direction, that is to 
say $g(\vec H, \vec n)$= tr$\bm{K}_{\vec n}$, is termed ``expansion 
along $\vec n$'' of $S$. 

For a spacelike surface there are two {\em independent} normal vector fields, and thus they can be chosen to be null and future-pointing (say). Let them be $\vec{\ell}, \vec k \in \mathfrak{X}(S)^{\perp}$, and add the normalization condition $g(\vec{\ell},\vec{k})=-1$. There remains the freedom
\begin{equation}
\vec{\ell} \longrightarrow \vec{\ell}'=\sigma^2 \vec{\ell}, \hspace{1cm}
\vec{k} \longrightarrow \vec{k}'=\sigma^{-2} \vec{k} \label{free}
\end{equation}
where $\sigma^2$ is a positive function defined only on $S$. Thus, the shape tensor decomposes as
\be
\vec{\bm{K}}=-\bm{K}_{\vec k}\,\, \vec\ell -\bm{K}_{\vec\ell}\,\, \vec k \label{shape}
\ee
and the mean curvature vector as
\be
\vec{H}\equiv -(\mbox{tr}\, \bm{K}_{\vec k})\,\, \vec\ell - (\mbox{tr}\, \bm{K}_{\vec\ell})\,\, \vec k \, .\label{mean}
\ee
One can also define the ``determinant of $\vec{\bm{K}}$'', that is to say the vector field
\be
\vec G \equiv -(\det \bm{K}_{\vec k})\, \vec \ell - (\det \bm{K}_{\vec \ell})\, \vec k,
\hspace{1cm}  \vec G \in \mathfrak{X}(S)^{\perp}
\label{det} 
\ee
Both (\ref{shape}) and (\ref{mean})  are invariant under
transformations (\ref{free}), but {\em not} (\ref{det}). However, the norm of $\vec G$:
$$
g(\vec G,\vec G) =-2 \det \bm{K}_{\vec k}\det \bm{K}_{\vec \ell}=
-2\det \left(\bm{K}_{\vec k}\bm{K}_{\vec \ell}\right),
$$ 
and therefore its causal orientation too, are also invariant under (\ref{free}).

\section{Classification of either $\bm{K}_{\vec k}$ or $\bm{K}_{\vec \ell}$}
Fix a point $x\in S$. Then either of $\bm{K}_{\vec k}$ or $\bm{K}_{\vec \ell}$ is a $2\times 2$ symmetric real matrix, and it can readily be algebraically classified: it is always diagonalizable (with respect to $\bm{\gamma}$) over $\mathbb{R}$.

Choose (say) $\bm{K}_{\vec k}|_{x}$, and let $\lambda_{1}$ and $\lambda_{2}$ be its eigenvalues, with $|\lambda_{1}|\geq |\lambda_{2}|$. 
There are eight possible types according to the signs of tr$\bm{K}_{\vec k}|_{x}$ and $\det \bm{K}_{\vec k}|_{x}$, plus some degenerate cases. The classification is shown in Table \ref{t1}. The notation is as follows: 
$$
\left(\mbox{sign($\lambda_{1}$)},\mbox{sign($\lambda_{2}$)}\right)
$$
where sign($\lambda_{i})\in\{+, 0, -\}$ ($i\in\{1,2\}$) according to whether $\lambda_{i}$ is positive, zero, or negative, respectively. Notice that the symbols are {\em ordered}, as the left entry is the one corresponding to the eigenvalue with greater magnitude (in absolute value). Thus, for instance, $(-,+)$ is the symbol for a $\bm{K}_{\vec k}|_{x}$ with eigenvalues of opposite signs and such that the negative one has greater magnitude than the positive one. If the two eigenvalues have the same magnitude, $|\lambda_{1}|= |\lambda_{2}|$, then only one symbol is written ---using $\pm$ for the case with different signs--- so that $(+)$, $(-)$, $(0)$ and $(\pm)$ correspond to the cases where the two eigenvalues are equal and positive, equal and negative, vanishing, or equal in magnitude with opposite signs, respectively.

Some standard terminology,, borrowed from General Relativity and Classical Differential Geometry, may be used for some of these types. Thus, the surface is said to be 
{\em $\vec k$-expanding}, {\em $\vec k$-contracting}, or {\em $\vec k$-stationary} at $x\in S$ according to the whether tr$\bm{K}_{\vec k}|_{x}$ is positive, negative, or zero, respectively. Also, $S$ is called {\em (strictly) future $\vec k$-convex} at $x\in S$ if  $\bm{K}_{\vec k}|_{x}$ is positive (definite) semi-definite (cases $(+,+)$ and $(+,0)$); and analogously for the cases $(-,-)$ and $(-,0)$ replacing future by past.

Regarding the degenerate cases with $\lambda_{1}=\lambda_{2}$ (that is,  $(+)$, $(-)$, $(0)$), the surface is said to be {\em $\vec k$-umbilical} at $x$ ---equivalently, $x$ is called a $\vec k$-umbilical point--- if $\bm{K}_{\vec k}|_{x}$ has type $(+)$ or $(-)$ there. These are characterized by the condition $(\mbox{tr}\bm{K}_{\vec k}|_{x})^2=4\det \bm{K}_{\vec k}|_{x}> 0$. They could also be called {\em $\vec k$-shear-free} points, because a hypersurface-orthogonal geodesic congruence tangent to $\vec k|_{x}$ at $x$ would be shear-free there. 
The extreme case with $\bm{K}_{\vec k}|_{x}= \bm{0}$, which in particular is
$\vec k$-stationary, corresponds to type $(0)$ and will be called a {\em $\vec k$-subgeodesic point}---see the beginning of section \ref{sec:primary} for a justification. Everything is analogous for the other null direction $\vec\ell$. 

\begin{table}[ht]
\begin{tabular}[c]{c||c|c|c||}
\thicklines{\put(37,-5){\line(-5,2){25}}\put(37,-4.5){\line(-5,2){25}}}
\begin{tabular}{c}
\hspace{14mm} tr$\bm{K}_{\vec k}|_{x}$\\
\hspace{-1cm}$\det\bm{K}_{\vec k}|_{x}$
\end{tabular} &
\begin{tabular}{c}
$> 0$\\
\scriptsize{($\vec{k}$-expanding)} 
\end{tabular}  
& \begin{tabular}{c}
$= 0$\\
\scriptsize{($\vec{k}$-stationary)} 
\end{tabular} 
& \begin{tabular}{c}
$< 0$\\
\scriptsize{($\vec{k}$-contracting)} 
\end{tabular}  \\ 
\hline\hline
$>0$ & $(+,+)\supset (+)$ & \raisebox{-1ex}{Not possible} & $(-,-)\supset (-)$ \\
\scriptsize{($\bm{K}_{\vec k}|_{x}$ is definite)} & \scriptsize{($\bm{K}_{\vec k}|_{x}$ is positive definite)} & & \scriptsize{($\bm{K}_{\vec k}|_{x}$ is negative definite)}\\
\hline 
$=0$ & $(+,0)$ & $(0)$ & $(-,0)$ \\
\scriptsize{($\bm{K}_{\vec k}|_{x}$ is semi-definite)} & 
\scriptsize{($\bm{K}_{\vec k}|_{x}$ is positive semi-definite)} &
\scriptsize{($\bm{K}_{\vec k}|_{x}$ vanishes)} & \scriptsize{($\bm{K}_{\vec k}|_{x}$ is negative semi-definite)}\\
\hline 
\begin{tabular}{c}
$<0$\\
\scriptsize{($\bm{K}_{\vec k}|_{x}$ is not definite)}
\end{tabular} & $(+,-)$ & $(\pm)$ & $(-,+)$\\
\hline\hline
\end{tabular}
\caption{\small The algebraic types of the matrix $\bm{K}_{\vec k}|_{x}$ at a point $x\in S$ according to the signs of its determinant and trace. For explanation of the symbols, see the main text.\label{t1}}
\end{table}

\section{Combined classification of $\bm{K}_{\vec k}$ and $\bm{K}_{\vec \ell}$}
The classification becomes more interesting and refined when the algebraic cases of the two null second fundamental forms $\bm{K}_{\vec k}$ and $\bm{K}_{\vec \ell}$ are taken together. This can be done in two steps: first, by just combining all possibilities; second, by taking also into account the relative orientations of the eigen-directions. 
\subsection{Algebraic combination}
The first basic step is to consider the different possibilities for the two traces. This is shown in Table \ref{t2}, which is symmetric. 
\begin{table}[ht]
\begin{tabular}[c]{c||c|c|c||}
\thicklines{\put(36,-5){\line(-5,2){25}}\put(36,-4.5){\line(-5,2){25}}}
\begin{tabular}{c}
\hspace{14mm} tr$\bm{K}_{\vec k}|_{x}$\\
\hspace{-1cm} tr$\bm{K}_{\vec \ell}|_{x}$
\end{tabular} &
\begin{tabular}{c}
$> 0$\\
\scriptsize{($\vec{k}$-expanding)} 
\end{tabular}  
& \begin{tabular}{c}
$= 0$\\
\scriptsize{($\vec{k}$-stationary)} 
\end{tabular} 
& \begin{tabular}{c}
$< 0$\\
\scriptsize{($\vec{k}$-contracting)} 
\end{tabular}  \\ 
\hline\hline
\begin{tabular}{c}
$>0$\\
\scriptsize{($\vec\ell$-expanding)}
\end{tabular}  & expanding & semi-expanding & mixed \\
\hline 
\begin{tabular}{c}
$=0$\\
\scriptsize{($\vec\ell$-stationary)}
\end{tabular}  & semi-expanding & stationary & semi-contracting \\
\hline 
\begin{tabular}{c}
$<0$\\
\scriptsize{($\vec\ell$-contracting)}
\end{tabular} & mixed & semi-contracting & contracting \\
\hline\hline
\end{tabular}
\caption{\small The possible combinations of the traces at a point $x\in S$. Each of these cases can be characterized by the causal character and orientation of the mean curvature vector $\vec H$, see the main text. Implicitly, one is thinking of time as flowing to the future, so that ``expanding'' means ``growing larger as time passes'', and analogously for ``contracting''.\label{t2}}
\end{table}
Each of the six different possibilites can be invariantly characterized by means of the causal orientation of the mean curvature vector as follows:
\begin{center}
\begin{tabular}{ccc}
Name & & $\vec H|_{x}$ \\
\hline
Expanding & $\Longleftrightarrow$ & past-pointing timelike\\
Semi-expanding & $\Longleftrightarrow$ & past-pointing null\\
Mixed & $\Longleftrightarrow$ & spacelike\\
Stationary & $\Longleftrightarrow$ & vanishes \\
Semi-contracting & $\Longleftrightarrow$ & future-pointing null\\
Contracting & $\Longleftrightarrow$ & future-pointing timelike\\
\end{tabular}
\end{center}

The full combination of the eight types (plus degeneracies) for each of 
$\bm{K}_{\vec k}|_{x}$ or $\bm{K}_{\vec \ell}|_{x}$ leads to the complete algebraic combined classification with a total of 64 cases (plus degeneracies) of which 9 are expanding, 12 semi-expanding, 18 mixed, 4 stationary, 12 semi-contracting, and 9 contracting.
This is shown in Table \ref{t3}. The notation to be used just mingles the types for both matrices separating them with a vertical bar, putting those corresponding to $\bm{K}_{\vec \ell}|_{x}$ on the left (say):
$$
\left(\mbox{sign($\mu_{1}$)},\mbox{sign($\mu_{2}$)}\right)|
\left(\mbox{sign($\lambda_{1}$)},\mbox{sign($\lambda_{2}$)}\right)
$$
where $\mu_{1}$ and $\mu_{2}$ stand for the eigenvalues of $\bm{K}_{\vec \ell}|_{x}$, such that $|\mu_{1}|\geq |\mu_{2}|$. Thus, for instance $(+,0)|(+,-)$ means that $\bm{K}_{\vec\ell}|_{x}$ has one positive and one vanishing eigenvalue, while $\bm{K}_{\vec k}|_{x}$ has two eigenvalues of opposite signs, the positive one with larger magnitude. This particular case is {\em expanding} (both traces are positive, so that $\vec H|_{x}$ is past timelike), 
\newpage
\changepage{3cm}{}{}{}{}{-1cm}{}{}{}
\thispagestyle{empty}
\begin{landscape}
\begin{table}[th]
\footnotesize{
\begin{tabular}[c]{l|c|||c|c|c||c|c||c|c|c|||}
\multicolumn{2}{r|||}{\thicklines{\put(-2,-5){\line(-3,2){10}}\put(-2,-4.3){\line(-3,2){10}}}
$\bm{K}_{\vec k}|_{x}$} & \multicolumn{3}{|c||}{$\vec k$-expanding} &
\multicolumn{2}{|c||}{$\vec k$-stationary} &
\multicolumn{3}{|c|||}{$\vec k$-contracting} \\
\cline{3-10}
\multicolumn{2}{l|||}{\thicklines{\put(35,-5){\line(-3,2){13}}\put(35,-4.3){\line(-3,2){13}}}
$\bm{K}_{\vec \ell}|_{x}$} & $(+,+)$ \raisebox{-1ex}{\scriptsize{$(+)$}} & $(+,0)$ & $(+,-)$ & $(0)$ 
& $(\pm)$ & $(-,+)$ & $(-,0)$ & $(-,-)$ \raisebox{-1ex}{\scriptsize{$(-)$}} \\
\hline\hline\hline
$\vec\ell$--ex-& $(+,+)$ \raisebox{-1ex}{\scriptsize{$(+)$}}  & 
\begin{tabular}{c}
$(+,+)|(+,+)$ \\
\tiny{$(+)|(+,+)$}\\ \tiny{$(+,+)|(+)$}\\\tiny{$(+)|(+)$}
\end{tabular} & 
\begin{tabular}{c}
$(+,+)|(+,0)$ \\
\tiny{$(+)|(+,0)$}
\end{tabular} &
\begin{tabular}{c}
$(+,+)|(+,-)$ \\
\tiny{$(+)|(+,-)$}
\end{tabular} &
\begin{tabular}{c}
$(+,+)|(0)$ \\
\tiny{$(+)|(0)$}
\end{tabular} &
\begin{tabular}{c}
$(+,+)|(\pm)$ \\
\tiny{$(+)|(\pm)$}
\end{tabular} &
\begin{tabular}{c}
$(+,+)|(-,+)$ \\
\tiny{$(+)|(-,+)$}
\end{tabular} &
\begin{tabular}{c}
$(+,+)|(-,0)$ \\
\tiny{$(+)|(-,0)$}
\end{tabular} &
\begin{tabular}{c}
$(+,+)|(-,-)$ \\
\tiny{$(+)|(-,-)$}\\ \tiny{$(+,+)|(-)$}\\\tiny{$(+)|(-)$}
\end{tabular} \\
\cline{2-10}
pan-& $(+,0)$ &
\begin{tabular}{c}
$(+,0)|(+,+)$ \\
\tiny{$(+,0)|(+)$}
\end{tabular} &
$(+,0)|(+,0)$ & $(+,0)|(+,-)$ & $(+,0)|(0)$ & $(+,0)|(\pm)$ & $(+,0)|(-,+)$ & $(+,0)|(-,0)$ &
\begin{tabular}{c}
$(+,0)|(-,-)$ \\
\tiny{$(+,0)|(-)$}
\end{tabular}\\
\cline{2-10}
ding & $(+,-)$ &
\begin{tabular}{c}
$(+,-)|(+,+)$ \\
\tiny{$(+,-)|(+)$}
\end{tabular} &
$(+,-)|(+,0)$ & $(+,-)|(+,-)$ & $(+,-)|(0)$ & $(+,-)|(\pm)$ & $(+,-)|(-,+)$ & $(+,-)|(-,0)$ &
\begin{tabular}{c}
$(+,-)|(-,-)$ \\
\tiny{$(+,-)|(-)$}
\end{tabular}\\
\hline\hline
$\vec\ell$--stati- & $(0)$ &
\begin{tabular}{c}
$(0)|(+,+)$ \\
\tiny{$(0)|(+)$}
\end{tabular} &
$(0)|(+,0)$ & $(0)|(+,-)$ & $(0)|(0)$ & $(0)|(\pm)$ & $(0)|(-,+)$ & $(0)|(-,0)$ &
\begin{tabular}{c}
$(0)|(-,-)$ \\
\tiny{$(0)|(-)$}
\end{tabular}\\
\cline{2-10}
onary & $(\pm)$ &
\begin{tabular}{c}
$(\pm)|(+,+)$ \\
\tiny{$(\pm)|(+)$}
\end{tabular} &
$(\pm)|(+,0)$ & $(\pm)|(+,-)$ & $(\pm)|(0)$ & $(\pm)|(\pm)$ & $(\pm)|(-,+)$ & $(\pm)|(-,0)$ &
\begin{tabular}{c}
$(\pm)|(-,-)$ \\
\tiny{$(\pm)|(-)$}
\end{tabular}\\
\hline\hline
$\vec\ell$--con- & $(-,+)$ &
\begin{tabular}{c}
$(-,+)|(+,+)$ \\
\tiny{$(-,+)|(+)$}
\end{tabular} &
$(-,+)|(+,0)$ & $(-,+)|(+,-)$ & $(-,+)|(0)$ & $(-,+)|(\pm)$ & $(-,+)|(-,+)$ & $(-,+)|(-,0)$ &
\begin{tabular}{c}
$(-,+)|(-,-)$ \\
\tiny{$(-,+)|(-)$}
\end{tabular}\\
\cline{2-10}
trac- & $(-,0)$ &
\begin{tabular}{c}
$(-,0)|(+,+)$ \\
\tiny{$(-,0)|(+)$}
\end{tabular} &
$(-,0)|(+,0)$ & $(-,0)|(+,-)$ & $(-,0)|(0)$ & $(-,0)|(\pm)$ & $(-,0)|(-,+)$ & $(-,0)|(-,0)$ &
\begin{tabular}{c}
$(-,0)|(-,-)$ \\
\tiny{$(-,0)|(-)$}
\end{tabular}\\
\cline{2-10}
ting & $(-,-)$ \raisebox{-1ex}{\scriptsize{$(-)$}} &
\begin{tabular}{c}
$(-,-)|(+,+)$ \\
\tiny{$(-)|(+,+)$}\\ \tiny{$(-,-)|(+)$}\\\tiny{$(-)|(+)$}
\end{tabular} & 
\begin{tabular}{c}
$(-,-)|(+,0)$ \\
\tiny{$(-)|(+,0)$}
\end{tabular} &
\begin{tabular}{c}
$(-,-)|(+,-)$ \\
\tiny{$(-)|(+,-)$}
\end{tabular} &
\begin{tabular}{c}
$(-,-)|(0)$ \\
\tiny{$(-)|(0)$}
\end{tabular} &
\begin{tabular}{c}
$(-,-)|(\pm)$ \\
\tiny{$(-)|(\pm)$}
\end{tabular} &
\begin{tabular}{c}
$(-,-)|(-,+)$ \\
\tiny{$(-)|(-,+)$}
\end{tabular} &
\begin{tabular}{c}
$(-,-)|(-,0)$ \\
\tiny{$(-)|(-,0)$}
\end{tabular} &
\begin{tabular}{c}
$(-,-)|(-,-)$ \\
\tiny{$(-)|(-,-)$}\\ \tiny{$(-,-)|(-)$}\\\tiny{$(-)|(-)$}
\end{tabular} \\
\hline\hline\hline
\end{tabular}
}
\caption{\small This is a ``magnification" of Table \ref{t2} providing the 64 algebraic cases ---plus degeneracies, which are shown in smaller letters--- at a point $x\in S$. The nine possibilities (six due to the $\vec{\ell} \leftrightarrow \vec{k}$ symmetry) in Table \ref{t2} are separated here by double lines. The notation is explained in the main text.\label{t3}}
\end{table}
\end{landscape}
\noindent
and has a null vector $\vec G|_x$. If one (or both) of the types is degenerate, then the symbol is accordingly written, e.g.\ $(+)|(0,-)$ or $(+)|(-)$.

\changepage{-3cm}{}{}{}{}{}{}{}{}
Therefore, the 64 cases in Table \ref{t3} can be grouped in 9 classes corresponding to the 9 cases shown in Table \ref{t2}. This has been incorporated into Table \ref{t3} by means of double-line separations. Observe, however, that due to the symmetry in the interchange of $\vec \ell$ with $\vec k$ there are several cases which have the same name and the same invariant characterization. For instance, $(+,+)|(0)$ and $(0)|(+,+)$ which are defined by past-pointing {\em co-linear} and null $\vec G|_x$ and $\vec H|_x$ ---see Table \ref{t4}---. This even happens within the 9 boxes defined by Table \ref{t2} when maginified according to Table \ref{t3}. An example of this is provided by the two cases  $(+,+),(+,0)$ and $(+,0),(+,+)$, characterized by past-pointing timelike $\vec H|_x$ and null $\vec G|_x$. 

Nevertheless, this $\vec k \leftrightarrow \vec\ell$ symmetry is broken in many practical cases, especially those of greater physical interest. Hence, whenever one can invariantly identify the two different null normals $\vec\ell$ and $\vec k$ ---such as in asymptotically flat cases, or for closed surfaces $S$, or in general whenever an inner and an outer part can be clearly identified---then each one of the 64 cases in Table \ref{t3} (but not its degeneracies) can be completely characterized by the causal character and orientation of both $\vec G|_x$ and $\vec H|_x$ .

\subsection{Relative orientation}
\label{relative}
Even though both $\bm{K}_{\vec\ell}|_x$ and $\bm{K}_{\vec k}|_x$ are diagonalizable, they may {\em not} be so simultaneously. This introduces a new parameter, which can be taken to be the relative orientation of the respective eigenbases. The value of this parameter can be used to refine the previous classification, and has some relevant consequences in physical applications.

The question is whether or not $\bm{K}_{\vec\ell}|_x$ and $\bm{K}_{\vec k}|_x$ commute, that is, if its commutator\footnote{In purity, one should only talk about the commutator of the endomorphisms $\hat{\bm{K}}_{\vec\ell}|_x$, $\hat{\bm{K}}_{\vec k}|_x\, : T_{x}S \rightarrow T_{x}S$ associated naturally to $\bm{K}_{\vec\ell}|_x$ and $\bm{K}_{\vec k}|_x$. Therefore, the notation $\left[\bm{K}_{\vec k}|_x,\bm{K}_{\vec\ell}|_x\right]$ used in the main text must be understood as the 2-form associated naturally to the usual commutator $\left[\hat{\bm{K}}_{\vec k}|_x,\hat{\bm{K}}_{\vec\ell}|_x\right]$ by ``lowering'' its upper index with $\bm{\gamma}$.}
$$
\left[\bm{K}_{\vec k}|_x,\bm{K}_{\vec\ell}|_x\right]
$$
vanishes or not. Let $\{\vec{e}_1,\vec{e_2}\}$ be the orthonormal eigenbasis of $\bm{K}_{\vec k}|_x$ corresponding to the eigenvalues $\lambda_1,\lambda_2$. Similarly, let $\{\vec{E}_1,\vec{E}_2\}$ be the orthonormal eigenbasis of $\bm{K}_{\vec\ell}|_x$ with eigenvalues $\mu_1,\mu_2$, respectively. Recall that $|\mu_{1}|\geq |\mu_{2}|$ and $|\lambda_{1}|\geq |\lambda_{2}|$. It is an exercise to check that, in any one of those two bases,
\bean
\left[\bm{K}_{\vec k}|_x,\bm{K}_{\vec\ell}|_x\right]=\bm{K}_{\vec\ell}|_x\left(\vec{e}_1,\vec{e}_2\right)
\left(\lambda_1-\lambda_2\right)\left(\begin{array}{cc}
0 & 1\\
-1 & 0
\end{array}
\right)\\
=-\bm{K}_{\vec k}|_x (\vec{E}_1,\vec{E}_2)
\left(\mu_1-\mu_2\right)\left(\begin{array}{cc}
0 & 1\\
-1 & 0
\end{array}
\right)
\eean
so that in general
$$
\bm{K}_{\vec\ell}|_x\left(\vec{e}_1,\vec{e}_2\right)
\left(\lambda_1-\lambda_2\right)+\bm{K}_{\vec k}|_x (\vec{E}_1,\vec{E}_2)
\left(\mu_1-\mu_2\right)=0.
$$
Of course, the orthonormal eigenbasis $\{\vec{e}_1,\vec{e}_2\}$ is not fixed in the degenerate cases $(+)$, $(0)$ and $(-)$, because in those cases the matrix $\bm{K}_{\vec k}|_x$ is proportional to the identity and  
\begin{landscape}
\begin{table}[th]
\begin{tabular}[c]{l|c|||c|c|c||c|c||c|c|c|||}
\multicolumn{2}{c|||}{Causal character} & \multicolumn{3}{|c||}{$\vec k$-expanding} &
\multicolumn{2}{|c||}{$\vec k$-stationary} &
\multicolumn{3}{|c|||}{$\vec k$-contracting} \\
\cline{3-10}
\multicolumn{2}{c|||}{of $\vec G|_x$} & $(+,+)$ \raisebox{-1ex}{\scriptsize{$(+)$}} & $(+,0)$ & $(+,-)$ & $(0)$ 
& $(\pm)$ & $(-,+)$ & $(-,0)$ & $(-,-)$ \raisebox{-1ex}{\scriptsize{$(-)$}} \\
\hline\hline\hline
$\vec\ell$--ex-& $(+,+)$ \raisebox{-1ex}{\scriptsize{$(+)$}}  & 
\begin{tabular}{c}
past \\ timelike
\end{tabular}& 
\begin{tabular}{c}
past \\
null
\end{tabular} &
spacelike &
\begin{tabular}{c}
past \\
null
\end{tabular} &
spacelike &
spacelike &
\begin{tabular}{c}
past \\
null
\end{tabular} &
\begin{tabular}{c}
past \\ timelike
\end{tabular}\\
\cline{2-10}
pan-& $(+,0)$ &
\begin{tabular}{c}
past \\
null
\end{tabular}  &
zero & 
\begin{tabular}{c}
future \\
null
\end{tabular}  & zero &
\begin{tabular}{c}
future \\
null
\end{tabular}  & 
\begin{tabular}{c}
future \\
null
\end{tabular}  & zero &
\begin{tabular}{c}
past \\
null
\end{tabular} \\
\cline{2-10}
ding & $(+,-)$ &
spacelike &
\begin{tabular}{c}
future \\
null
\end{tabular} & 
\begin{tabular}{c}
future \\
timelike
\end{tabular} & 
\begin{tabular}{c}
future \\
null
\end{tabular} & 
\begin{tabular}{c}
future \\
timelike
\end{tabular} & 
\begin{tabular}{c}
future \\
timelike
\end{tabular} & 
\begin{tabular}{c}
future \\
null
\end{tabular} &
spacelike\\
\hline\hline
$\vec\ell$--stati- & $(0)$ &
\begin{tabular}{c}
past \\
null
\end{tabular}  &
zero & 
\begin{tabular}{c}
future \\
null
\end{tabular} & zero & 
\begin{tabular}{c}
future \\
null
\end{tabular} &
\begin{tabular}{c}
future \\
null
\end{tabular} & zero &
\begin{tabular}{c}
past \\
null
\end{tabular}\\
\cline{2-10}
onary & $(\pm)$ &
spacelike &
\begin{tabular}{c}
future \\
null
\end{tabular} & 
\begin{tabular}{c}
future \\
timelike
\end{tabular} & 
\begin{tabular}{c}
future \\
null
\end{tabular} & 
\begin{tabular}{c}
future \\
timelike
\end{tabular} & 
\begin{tabular}{c}
future \\
timelike
\end{tabular} & 
\begin{tabular}{c}
future \\
null
\end{tabular} &
spacelike \\
\hline\hline
$\vec\ell$--con- & $(-,+)$ &
spacelike &
\begin{tabular}{c}
future \\
null
\end{tabular} &
\begin{tabular}{c}
future \\
timelike
\end{tabular} & 
\begin{tabular}{c}
future \\
null
\end{tabular} & 
\begin{tabular}{c}
future \\
timelike
\end{tabular} & 
\begin{tabular}{c}
future \\
timelike
\end{tabular} & 
\begin{tabular}{c}
future \\
null
\end{tabular} &
spacelike \\
\cline{2-10}
trac- & $(-,0)$ &
\begin{tabular}{c}
past \\
null
\end{tabular} &
zero & 
\begin{tabular}{c}
future \\
null
\end{tabular} & zero & 
\begin{tabular}{c}
future \\
null
\end{tabular} & 
\begin{tabular}{c}
future \\
null
\end{tabular} & zero &
\begin{tabular}{c}
past \\
null
\end{tabular}\\
\cline{2-10}
ting & $(-,-)$ \raisebox{-1ex}{\scriptsize{$(-)$}} &
\begin{tabular}{c}
past \\
timelike
\end{tabular} & 
\begin{tabular}{c}
past \\
null
\end{tabular} &
spacelike &
\begin{tabular}{c}
past \\
null
\end{tabular} &
spacelike &
spacelike &
\begin{tabular}{c}
past \\
null
\end{tabular} &
\begin{tabular}{c}
past \\
timelike
\end{tabular} \\
\hline\hline\hline
\end{tabular}
\caption{\small The causal character and orientation of the vector $\vec G|_x$ for each of the cases in Table \ref{t3}. The resulting table is symmetric. However, if one can distinguish $\vec\ell$ from $\vec k$ in an invariant way, then this symmetry is broken and each of the 64 cases can be identified by the causal character of both $\vec H|_x$ and $\vec G|_x$, see the main text.\label{t4}}
\end{table}
\end{landscape}
\noindent
{\em any} possible orthonormal basis will do. Analogously for the degenerate cases of $\bm{K}_{\vec\ell}|_x$. Nevertheless, 
if any of the two matrices belongs to one of the degenerate types, obviously they do commute, hence one can select the orthonormal eigenbasis of the non-degenerate one if this exists, or any orthonormal basis if both matrices are degenerate at $x$.

Still, there remains the issue that, for the $(\pm)$ type, one cannot distinguish between the eigenbases $\{\vec{e}_1,\vec{e_2}\}$ and $\{\vec{e}_2,\vec{e}_1\}$; this problem is fixed by adopting the convention that, {\em for the type $(\pm)$}, the positive eigenvalue corresponds to the first eigenvector $\vec{e}_1$. 

Keeping this in mind, consider the generic situation in which neither $\bm{K}_{\vec k}|_x$ nor $\bm{K}_{\vec\ell}|_x$ has a degenerate type. Without loss of generality one can take the respective orthonormal eigenbases having the same orientation, $\bm{e}_1\wedge\bm{e}_2 =\bm{E}_1\wedge\bm{E}_2$, and with this choice they are related by means of a rotation with angle $\alpha \in (-\pi/2 ,\pi/2]$
\bean
\vec{E}_1=\cos\alpha\, \vec{e}_1 +\sin\alpha\,\vec{e}_2 \, ,\\
\vec{E}_2=-\sin\alpha\, \vec{e}_1+\cos\alpha\, \vec{e}_2\, .
\eean

\noindent
Then a trivial calculation leads to 
$$
\sin 2\alpha =-\, \frac{2\bm{K}_{\vec k}|_x(\vec{E}_1,\vec{E}_2)}{\lambda_1-\lambda_2}=
\frac{2\bm{K}_{\vec\ell}|_x\left(\vec{e}_1,\vec{e}_2\right)}{\mu_1-\mu_2} 
$$
or equivalently, given the adopted conventions for the eigenvalues,
$$
\sin 2\alpha =-\, \frac{2\bm{K}_{\vec k}|_x(\vec{E}_1,\vec{E}_2)}{\epsilon_{\vec k}
\sqrt{(\mbox{tr}\bm{K}_{\vec k}|_x)^2-4\det\bm{K}_{\vec k}|_x}}=
\frac{2\bm{K}_{\vec\ell}|_x\left(\vec{e}_1,\vec{e}_2\right)}{\epsilon_{\vec\ell}
\sqrt{(\mbox{tr}\bm{K}_{\vec\ell}|_x)^2-4\det\bm{K}_{\vec\ell}|_x}} 
$$
where $\epsilon_{\vec k}$ and $\epsilon_{\vec\ell}$ are the signs of the traces of $\bm{K}_{\vec k}|_x$ and $\bm{K}_{\vec\ell}|_x$ respectively. Therefore, one can also write
\be
\left[\bm{K}_{\vec k}|_x,\bm{K}_{\vec\ell}|_x\right]=\sin 2\alpha \, (\mu_1-\mu_2)
\left(\lambda_1-\lambda_2\right) \bm{e}_1\wedge\bm{e}_2
\label{comm}
\ee

The new parameter to be used in the classification is simply $\alpha$. One can add its value as a subscript to the different types, using the notation
$$
\left(\mbox{sign($\mu_{1}$)},\mbox{sign($\mu_{2}$)}\right)|
\left(\mbox{sign($\lambda_{1}$)},\mbox{sign($\lambda_{2}$)}\right)_{\alpha}\, .
$$
For instance, type $(+,+)|(+,+)_{\pi/3}$ is the expanding type, with past-pointing timelike vectors $\vec H|_x$ and $\vec G|_x$, and such that the ordered eigenbasis of $\bm{K}_{\vec\ell}|_x$ is rotated 60 degrees clockwise with respect to the ordered eigenbasis of $\bm{K}_{\vec k}|_x$.

Two particular values of $\alpha$ are distinguished: $\alpha =0$ and $\alpha =\pi/2$. In both cases the commutator (\ref{comm}) vanishes, so that the matrices $\bm{K}_{\vec k}|_x$ and $\bm{K}_{\vec\ell}|_x$ are simultaneously diagonalizable. The former case will be called {\em congruent}, and the latter {\em orthogonal}, so that one may say for example that the surface $S$ is ``congruently expanding" at any $x$ with type $(+,+)|(+,+)_{0}$, or that is ``orthogonally semi-contracting" at any $x$ with type $(\pm)|(-,0)_{\pi/2}$ ---note that in this case the two negative eigenvalues, $\mu_2$ and $\lambda_1$, have parallel eigenvectors---.

For generic values of $\alpha$, one can ask whether or not there are directions $\vec v\in T_{x}S$ such that $\vec{\bm{K}}|_x(\vec v,\vec v)$ is causal (timelike or null), which leads to the resolution of when
$\bm{K}_{\vec k}|_x$ and $\bm{K}_{\vec\ell}|_x$ are positive or negative definite over a
{\em common} set of directions. The non-trivial cases are defined by non-definite $\bm{K}_{\vec k}|_x$ or $\bm{K}_{\vec\ell}|_x$, that is, when either $\det \bm{K}_{\vec k}|_x$ or $\det \bm{K}_{\vec\ell}|_x$ is negative. Assume then that (say)
$\det \bm{K}_{\vec k}|_x < 0$, so that $\lambda_1\lambda_2 <0$. Then, $\bm{K}_{\vec k}|_x$ is positive (respectively negative) definite over the set of directions $\vec v\in TS$ defined by
\be
\vec v =\cos \beta \, \vec{e}_1 +\sin\beta\, \vec{e}_2 , \hspace{1cm} (-\pi/2<\beta\leq \pi/2) \label{v}
\ee
with
$$
\sin^2\beta < \frac{\lambda_1}{\lambda_1 -\lambda_2} 
\equiv B_{c}\hspace{2mm} \mbox{[case $(+,-)$]}, \hspace{9mm} \sin^2\beta > B_{c}\hspace{2mm} \mbox{[case $(-,+)$]}
$$
(or respectively
$$
\sin^2\beta > B_{c} \hspace{2mm} \mbox{[case $(+,-)$]}, \hspace{9mm} \sin^2\beta < B_{c}\hspace{2mm} \mbox{[case $(-,+)$]}\, )
$$
and $\bm{K}_{\vec k}|_x(\vec v,\vec v)=0$ at the boundaries defined by the critical values: $\sin^2\beta = B_{c}$. Observe that $B_{c}$ can take the following explicit values
$$
B_{c}=\frac{\lambda_{1}}{\lambda_{1}-\lambda_{2}}\left\{\begin{array}{ccc}
\in (1/2,1) &  \mbox{for types} & (+,-)\,\, \mbox{or}\,\, (-,+) \\
\\
\displaystyle{=\frac{1}{2}} &  \mbox{for type} & (\pm) \\
\end{array}
\right.
$$
Another pertinent question is whether or not there are {\em orthonormal} bases $\{\vec{v}_{1},\vec{v}_{2}\}$ such that both $\bm{K}_{\vec k}|_x(\vec{v}_{i},\vec{v}_{i})$ ($i=1,2$) have the same sign. Letting $\vec{v}_{1}$ take the form (\ref{v}) so that $\vec{v}_{2}=-\sin\beta\, \vec{e}_1 + \cos \beta \, \vec{e}_2$, both of those values will be simultaneously positive for the case $(+,-)$ if
$$
B'_{c}< \sin^2\beta < B_{c}, \hspace{1cm} 
B'_{c}\equiv 1-B_{c}=\frac{\lambda_{2}}{\lambda_{2}-\lambda_{1}}
$$
and negative if 
$$
B'_{c}>  \sin^2\beta > B_{c}\, .
$$
The case $(-,+)$ is analogous by reversing signs.
Therefore, for the case $(+,-)$, orthonormal bases $\{\vec{v}_{1},\vec{v}_{2}\}$ such that both $\bm{K}_{\vec k}|_x(\vec{v}_{i},\vec{v}_{i})$ ($i=1,2$) are negative are forbidden; the other (positive) case is allowed. Similarly, orthonormal bases $\{\vec{v}_{1},\vec{v}_{2}\}$ such that both 
$\bm{K}_{\vec k}|_x(\vec{v}_{i},\vec{v}_{i})$ ($i=1,2$) are positive are forbidden in case $(-,+)$.

Analogous formulas, replacing $\lambda's$ by $\mu's$, can be given for the non-definite $\bm{K}_{\vec\ell}|_{x}$, and in this case the critical value (analogous to $B_{c}$) will be called $C_{c}$.

All of the above allows us to give a refined classification, based on the six cases of Table \ref{t2}, taking into account the value of $\alpha$ and the particular algebraic types of $\bm{K}_{\vec k}|_x$ and $\bm{K}_{\vec\ell}|_x$. This leads to many possibilities which, for the sake of clarity and readibility of the paper, are placed and carefully considered in the Appendix. A summary of this, paying attention to the permitted causal orientation of the vectors $\vec{\bm{K}}|_x(\vec v,\vec v)$ for $\vec v \in T_x S$, is presented in what follows. The summary is made visual by means of a set of Tables where the allowed causal orientations of $\vec{\bm{K}}|_x(\vec v,\vec v)$ are displayed for each case. The notation to be used is
\begin{center}
\begin{tabular}{ccc}
Symbol & & Causal orientation \\
\hline
$\downarrow$ & $\Longleftrightarrow$ & past-pointing timelike\\\\
$\swarrow$ or $\searrow$ & $\Longleftrightarrow$ & past-pointing null ($\propto \vec \ell$ or $\vec k$) \\\\
$\leftarrow$\, or\, $\rightarrow$ & $\Longleftrightarrow$ & spacelike\\\\
$0$ & $\Longleftrightarrow$ & vanishes \\\\
$\nearrow$ or $\nwarrow$ & $\Longleftrightarrow$ & future-pointing null ($\propto \vec \ell$ or $\vec k$)\\\\
$\uparrow$ & $\Longleftrightarrow$ & future-pointing timelike\\
\end{tabular}
\end{center}
Furthermore, the {\em dominant} orientation is highlighted with a larger arrow. This dominant orientation is defined as follows: take the values of $\beta$ in (\ref{v}) such that $\vec{\bm{K}}|_x(\vec v,\vec v)$ has a particular causal orientation. These values belong to sub-intervals, or particular points, in $(-\pi/2,\pi/2]$. The intervals may be disconnected. The total standard length of these intervals for the chosen orientation is a measure of its frequency. Thus, the dominant orientation is the one with the largest such measure if this exists (observe that sometimes there may be two orientations with the same frequency, so that none of them dominates.)

\subsubsection{Expanding points}
This is defined by a past-pointing mean curvature vector $\vec H|_{x}=$ tr$\vec{\bm{K}}|_{x}$, hence the ``mean tendency'' is that $\vec{\bm{K}}|_{x}(\vec v,\vec v)$ be past timelike for generic $\vec v \in T_{x}S$. Nevertheless, this averaged tendency is not exact and there arise several interesting situations, even with {\em future}-pointing $\vec{\bm{K}}|_{x}(\vec v,\vec v)$ for some $\vec v$, according to the subclassification in the 9 classes ---supplemented with the particular value of $\alpha$--- appearing at the left upper box of Table \ref{t3}. This leads to the many different cases considered in the Appendix. The summary of all this is given in the Table \ref{t5}.

\begin{table}
\begin{center}
\begin{tabular}{||c|c||}
Expanding $x\in S$ & 
Orientation of $\vec{\bm{K}}|_{x}(\vec v,\vec v)$ \\
\hline\hline
$(+,+)|(+,+)_{\alpha}$ & {\Huge $\downarrow$}  \\
\hline 
$(+,+)|(+,0)_{\alpha}$  &  {\Huge $\downarrow$}  $\searrow$\\
\hline
$(+,0)|(+,+)_{\alpha}$ &  {\Huge $\downarrow$} $\swarrow$ \\
\hline
$(+,0)|(+,0)_{\alpha\neq 0}$ &   {\Huge $\downarrow$} $\swarrow$ $\searrow$ \\
$(+,0)|(+,0)_{0}$ & {\Huge $\downarrow$} 0\\
\hline 
$(+,+)|(+,-)_{\alpha}$ &   {\Huge $\downarrow$} $\searrow$ $\rightarrow$ \\
\hline
$(+,-)|(+,+)_{\alpha}$ &   {\Huge $\downarrow$} $\swarrow$ $\leftarrow$ \\
\hline
$(+,0)|(+,-)_{|\alpha|< \alpha_c}$ &  {\Huge $\downarrow$} $\searrow$ $\rightarrow$ $\nearrow$ \\
$(+,0)|(+,-)_{|\alpha|> \alpha_c}$ &  {\Huge $\downarrow$}$\searrow$ $\rightarrow$ $\swarrow$ \\
$(+,0)|(+,-)_{|\alpha|= \alpha_c}$ &  {\Huge $\downarrow$} $\searrow$ $\rightarrow$ $0$ \\
\hline
$(+,-)|(+,0)_{|\alpha|< \alpha'_c}$ &  {\Huge $\downarrow$} $\swarrow$ $\leftarrow$ $\nwarrow$ \\
$(+,-)|(+,0)_{|\alpha|> \alpha'_c}$ &  {\Huge $\downarrow$}$\swarrow$ $\leftarrow$ $\searrow$ \\
$(+,-)|(+,0)_{|\alpha|= \alpha'_c}$ &  {\Huge $\downarrow$} $\swarrow$ $\leftarrow$ $0$ \\
\hline
$(+,-)|(+,-)_{|\alpha|< |\alpha'_c-\alpha_c|}$ &   {\Huge $\downarrow$} $\uparrow$ $\swarrow$ $\searrow$ $\rightarrow$ $\nwarrow$ $\nearrow$ \\
$(+,-)|(+,-)_{|\alpha|= \alpha'_c-\alpha_c>0}$ & {\Huge $\downarrow$} $\uparrow$  $\searrow$ $\rightarrow$ $\nearrow$ $0$ \\
$(+,-)|(+,-)_{|\alpha|= \alpha_c-\alpha'_c> 0}$ & {\Huge $\downarrow$} $\uparrow$ $\swarrow$ $\leftarrow$ $\nwarrow$ $0$ \\
$(+,-)|(+,-)_{|\alpha|= |\alpha'_c-\alpha_c|= 0}$ &   {\Huge $\downarrow$}  $\uparrow$ $0$\\
$(+,-)|(+,-)_{|\alpha'_c-\alpha_c|< |\alpha|< \alpha'_c+\alpha_c}$ & 
({\Huge $\downarrow$}) $\uparrow$ $\swarrow$ $\searrow$  
({\Huge $\leftarrow$}) ({\Huge $\rightarrow$}) $\nwarrow$ $\nearrow$ \\
$(+,-)|(+,-)_{|\alpha|= \alpha'_c+\alpha_c}$ & ({\Huge $\downarrow$}) $\swarrow$ $\searrow$  
({\Huge $\leftarrow$}) ({\Huge $\rightarrow$})  $0$\\
$(+,-)|(+,-)_{|\alpha|> \alpha'_c+\alpha_c}$ &  ({\Huge $\downarrow$}) $\swarrow$ $\searrow$  
({\Huge $\leftarrow$}) ({\Huge $\rightarrow$}) \\
\hline
\hline
\end{tabular}
\caption{\small The permitted causal orientations of $\vec{\bm{K}}|_{x}(\vec v,\vec v)$ for $\vec v\in T_x S$ for the {\em expanding} cases. The 9 possibilities appearing at the left upper box of Table \ref{t3} are represented here. The critical values for the orientation parameter $\alpha$ are given by 
$\alpha_c =\pi/2-\arcsin\sqrt{B_{c}}$ and
$\alpha'_c =\pi/2-\arcsin\sqrt{C_{c}}$. The larger arrows indicate the dominant causal orientation for the vectors $\vec{\bm{K}}|_{x}(\vec v,\vec v)$. See the main text, and specially the Appendix,  for further, more detailed, explanations. The special notation with round brackets used in the three lower files indicates that either of $\downarrow$, $\leftarrow$ or $\rightarrow$ may be the dominant orientation, depending on the concrete values of $\alpha$, $\lambda_{i}$ and $\mu_{i}$.   \label{t5}}
\end{center}
\end{table}

\subsubsection{Semi-Expanding points}
Now the mean tendency is that $\vec{\bm{K}}|_{x}(\vec v,\vec v)$ be past-pointing null; however, this is not the dominant case in general, as the different cases can be ``balanced" in order to produce that mean tendency. The full case is treated in the Appendix, and the summary is presented in Table \ref{t6}.

\begin{table}
\begin{center}
\begin{tabular}{||c|c||}
Semi-Expanding $x\in S$ & 
Orientation of $\vec{\bm{K}}|_{x}(\vec v,\vec v)$ \\
\hline\hline
$(+,+)|(0)$ & {\Huge $\searrow$}  \\
\hline 
$(+,0)|(0)$ &   {\Huge $\searrow$} $0$  \\
\hline 
$(+,-)|(0)$ &   {\Huge $\searrow$} $\nwarrow$ $0$ \\
\hline
$(+,+)|(\pm)_{\alpha}$  &  $\downarrow$  $\searrow$ $\rightarrow$\\
\hline
$(+,0)|(\pm)_{|\alpha|<\pi/4}$ &  {\Huge $\downarrow$} $\searrow$ $\rightarrow$ $\nearrow$ \\
$(+,0)|(\pm)_{|\alpha|>\pi/4}$ &   $\downarrow$ $\swarrow$ {\Huge $\rightarrow$}  $\searrow$ \\
$(+,0)|(\pm)_{|\alpha|=\pi/4}$ & $\downarrow$ $\searrow$ $\rightarrow$ $0$ \\
\hline
$(+,-)|(\pm)_{|\alpha|< |\alpha'_c-\pi/4|}$ &   {\Huge $\downarrow$} $\uparrow$ $\swarrow$ $\searrow$ $\rightarrow$ $\nwarrow$ $\nearrow$\\
$(+,-)|(\pm)_{|\alpha|=| \alpha'_c-\pi/4|}$ & {\Huge $\downarrow$} $\uparrow$  $\searrow$ $\rightarrow$ $\nearrow$ $0$ \\
$(+,-)|(\pm)_{|\alpha'_c-\pi/4|< |\alpha|< \alpha'_c+\pi/4}$ & ({\Huge $\downarrow$}) $\uparrow$ $\swarrow$ $\searrow$
({\Huge $\leftarrow$}) ({\Huge $\rightarrow$}) $\nwarrow$ $\nearrow$ \\
$(+,-)|(\pm)_{|\alpha|= \alpha'_c+\pi/4}$ & ({\Huge $\downarrow$}) $\swarrow$ $\searrow$
({\Huge $\leftarrow$}) ({\Huge $\rightarrow$}) $0$ \\
$(+,-)|(\pm)_{|\alpha|> \alpha'_c+\pi/4}$ & ({\Huge $\downarrow$}) $\swarrow$ $\searrow$
({\Huge $\leftarrow$}) ({\Huge $\rightarrow$})\\
\hline
\hline
\end{tabular}
\caption{\small The allowed causal orientations of $\vec{\bm{K}}|_{x}(\vec v,\vec v)$ with $\vec v\in T_x S$ for the {\em semi-expanding} cases. The 6 possibilities appearing at the central upper box of Table \ref{t3} are represented here. The critical value
$\alpha'_c$, as well as the notation with round brackets for the type 
$(+,-)|(\pm)_{|\alpha'_c-\pi/4|< |\alpha|< \alpha'_c+\pi/4}$, are the same as in Table \ref{t5}. . See the main text and Appendix for further details.  A similar table holds for the other semi-expanding case, corresponding to the left central box of Table \ref{t3}, by just interchanging the roles of $\vec k$ and $\vec \ell$. \label{t6}}
\end{center}
\end{table}

\begin{table}
\begin{center}
\begin{tabular}{||c|c||}
Stationary $x\in S$ & 
Orientation of $\vec{\bm{K}}|_{x}(\vec v,\vec v)$ \\
\hline\hline
$(0)|(0)$ & {\Huge $0$}  \\
\hline 
$(0)|(\pm)$  &  $\nearrow$  $\swarrow$ $0$\\
\hline 
$(\pm)|(0)$ & $\nwarrow$ $\searrow$ $0$  \\
\hline
$(\pm)|(\pm)_{0}$ &   $\downarrow$ $\uparrow$ $0$\\
$(\pm)|(\pm)_{0<|\alpha|\leq\pi/2}$ & $\downarrow$ $\uparrow$  $\searrow$ $\swarrow$ $\leftarrow$ $\rightarrow$ $\nearrow$ $\nwarrow$ \\
$(\pm)|(\pm)_{\pi/2}$ & $\leftarrow$ $\rightarrow$ $0$ \\
\hline
\hline
\end{tabular}
\caption{\small The allowed causal orientations of $\vec{\bm{K}}|_{x}(\vec v,\vec v)$ with $\vec v\in T_x S$ for the {\em stationary} cases. These are the 4 possibilities appearing at the central box of Table \ref{t3}. See the Appendix for further details. \label{t8}}
\end{center}
\end{table}

\begin{table}
\begin{center}
\begin{tabular}{||c|c||}
Mixed $x\in S$ & 
Orientation of $\vec{\bm{K}}|_{x}(\vec v,\vec v)$ \\
\hline\hline
$(-,-)|(+,+)_{\alpha}$ & {\Huge $\leftarrow$}  \\
\hline 
$(-,-)|(+,0)_{\alpha}$  &  $\nwarrow$  {\Huge $\leftarrow$}\\
\hline
$(-,0)|(+,+)_{\alpha}$ &   $\swarrow$ {\Huge $\leftarrow$} \\
\hline
$(-,0)|(+,0)_{\alpha\neq 0}$ &   $\swarrow$ $\nwarrow$  {\Huge $\leftarrow$}\\
$(-,0)|(+,0)_{0}$ & {\Huge $\leftarrow$} 0 \\
\hline 
$(-,-)|(+,-)_{\alpha}$ &   $\uparrow$ $\nwarrow$ {\Huge $\leftarrow$} \\
\hline
$(-,+)|(+,+)_{\alpha}$ & $\downarrow$ $\swarrow$   {\Huge $\leftarrow$} \\
\hline
$(-,0)|(+,-)_{|\alpha|< \alpha_c}$ &  $\uparrow$ $\nwarrow$  $\nearrow$  {\Huge $\leftarrow$} \\
$(-,0)|(+,-)_{|\alpha|> \alpha_c}$ &  $\uparrow$ $\nwarrow$ $\swarrow$ {\Huge $\leftarrow$} \\
$(-,0)|(+,-)_{|\alpha|= \alpha_c}$ & $\uparrow$ $\nwarrow$ {\Huge $\leftarrow$} $0$ \\
\hline
$(-,+)|(+,0)_{|\alpha|< \alpha'_c}$ & $\downarrow$ $\swarrow$  $\searrow$
 {\Huge $\leftarrow$} \\
$(-,+)|(+,0)_{|\alpha|> \alpha'_c}$ & $\downarrow$ $\swarrow$  $\nwarrow$
 {\Huge $\leftarrow$} \\
$(-,+)|(+,0)_{|\alpha|= \alpha'_c}$ & $\downarrow$ $\swarrow$ {\Huge $\leftarrow$}
$0$ \\
\hline
$(-,+)|(+,-)_{|\alpha|< |\alpha'_c-\alpha_c|}$ & $\uparrow$ $\swarrow$ $\searrow$ {\Huge $\leftarrow$} $\rightarrow$ $\nwarrow$ $\nearrow$\\
$(-,+)|(+,-)_{|\alpha|= \alpha'_c-\alpha_c>0}$ & $\uparrow$  $\nwarrow$ {\Huge $\leftarrow$} $\rightarrow$ $\nearrow$ $0$ \\
$(-,+)|(+,-)_{|\alpha|= \alpha_c-\alpha'_c> 0}$ & $\downarrow$ $\swarrow$ {\Huge $\leftarrow$} $\rightarrow$ $\searrow$ $0$\\
$(-,+)|(+,-)_{|\alpha|= |\alpha'_c-\alpha_c|= 0}$ &   {\Huge $\leftarrow$} $\rightarrow$ $0$\\
$(-,+)|(+,-)_{|\alpha'_c-\alpha_c|< |\alpha|< \alpha'_c+\alpha_c}$ &  ({\Huge $\uparrow$}) ({\Huge $\downarrow$}) $\swarrow$ $\nwarrow$ ({\Huge $\leftarrow$}) $\rightarrow$  $\searrow$ $\nearrow$ \\
$(-,+)|(+,-)_{|\alpha|= \alpha'_c+\alpha_c}$ & ({\Huge $\uparrow$}) ({\Huge $\downarrow$}) $\swarrow$ $\nwarrow$ ({\Huge $\leftarrow$}) $0$ \\
$(-,+)|(+,-)_{|\alpha|> \alpha'_c+\alpha_c}$ & ({\Huge $\uparrow$}) ({\Huge $\downarrow$}) $\swarrow$ $\nwarrow$ ({\Huge $\leftarrow$})\\
\hline
\hline
\end{tabular}
\caption{\small The possible causal orientations of $\vec{\bm{K}}|_{x}(\vec v,\vec v)$ for the {\em mixed} cases. The 9 possibilities appearing at the left lower box of Table \ref{t3} are represented here. The critical values of the orientation parameter $\alpha$ and the round-bracket notation are the same as in Table \ref{t5}. There is another similar table for the mixed cases appearing at the   right upper corner of Table \ref{t3}, by interchanging the roles of $\vec k$ and $\vec \ell$.\label{t7}}
\end{center}
\end{table}

\begin{table}
\begin{center}
\begin{tabular}{||c|c||}
Semi-Contracting $x\in S$ & 
Orientation of $\vec{\bm{K}}|_{x}(\vec v,\vec v)$ \\
\hline\hline
$(-,-)|(0)$ & {\Huge $\nwarrow$}  \\
\hline 
$(-,0)|(0)$ &   {\Huge $\nwarrow$} $0$  \\
\hline 
$(-,+)|(0)$ &   {\Huge $\nwarrow$} $\searrow$ $0$ \\
\hline
$(-,-)|(\pm)_{\alpha}$  &  $\uparrow$  $\nwarrow$ $\leftarrow$\\
\hline
$(-,0)|(\pm)_{|\alpha|<\pi/4}$ &  {\Huge $\uparrow$} $\nwarrow$ $\leftarrow$ $\swarrow$ \\
$(-,0)|(\pm)_{|\alpha|>\pi/4}$ &   $\uparrow$ $\nearrow$ {\Huge $\leftarrow$}  $\nwarrow$ \\
$(-,0)|(\pm)_{|\alpha|=\pi/4}$ & $\uparrow$ $\nwarrow$ $\leftarrow$ $0$ \\
\hline
$(-,+)|(\pm)_{|\alpha|< |\alpha'_c-\pi/4|}$ &   {\Huge $\uparrow$} $\downarrow$ $\nearrow$ $\nwarrow$ $\leftarrow$ $\searrow$ $\swarrow$\\
$(-,+)|(\pm)_{|\alpha|=| \alpha'_c-\pi/4|}$ & {\Huge $\uparrow$} $\downarrow$  $\nwarrow$ $\leftarrow$ $\swarrow$ $0$ \\
$(-,+)|(\pm)_{|\alpha'_c-\pi/4|< |\alpha|< \alpha'_c+\pi/4}$ & ( {\Huge $\uparrow$}) $\downarrow$ $\nearrow$ $\nwarrow$  
({\Huge $\leftarrow$}) ({\Huge $\rightarrow$}) $\searrow$ $\swarrow$ \\
$(-,+)|(\pm)_{|\alpha|= \alpha'_c+\pi/4}$ & ({\Huge $\uparrow$}) $\nearrow$ $\nwarrow$  
({\Huge $\leftarrow$}) ({\Huge $\rightarrow$}) $0$ \\
$(-,+)|(\pm)_{|\alpha|> \alpha'_c+\pi/4}$ & ({\Huge $\uparrow$}) $\nearrow$ $\nwarrow$  
({\Huge $\leftarrow$}) ({\Huge $\rightarrow$}) \\
\hline
\hline
\end{tabular}
\caption{\small The allowed causal orientations of $\vec{\bm{K}}|_{x}(\vec v,\vec v)$ with $\vec v\in T_x S$ for the {\em semi-contracting} types. The 6 possibilities appearing at the central lower box of Table \ref{t3} are represented here. The critical value
$\alpha'_c$, as well as the notation with round brackets are as in Table \ref{t6}. A similar table holds for the other semi-contracting case, corresponding to the right central box of Table \ref{t3}, by interchanging the roles of $\vec k$ and $\vec \ell$. \label{t9}}
\end{center}
\end{table}

\begin{table}
\begin{center}
\begin{tabular}{||c|c||}
Contracting $x\in S$ & 
Orientation of $\vec{\bm{K}}|_{x}(\vec v,\vec v)$ \\
\hline\hline
$(-,-)|(-,-)_{\alpha}$ & {\Huge $\uparrow$}  \\
\hline 
$(-,-)|(-,0)_{\alpha}$  &  {\Huge $\uparrow$}  $\nwarrow$\\
\hline
$(-,0)|(-,-)_{\alpha}$ &  {\Huge $\uparrow$} $\nearrow$ \\
\hline
$(-,0)|(-,0)_{\alpha\neq 0}$ &   {\Huge $\uparrow$} $\nearrow$ $\nwarrow$ \\
$(-,0)|(-,0)_{0}$ & {\Huge $\uparrow$} 0\\
\hline 
$(-,-)|(-,+)_{\alpha}$ &   {\Huge $\uparrow$} $\nwarrow$ $\rightarrow$ \\
\hline
$(-,+)|(-,-)_{\alpha}$ &   {\Huge $\uparrow$} $\nearrow$ $\leftarrow$ \\
\hline
$(-,0)|(-,+)_{|\alpha|< \alpha_c}$ &  {\Huge $\uparrow$} $\nwarrow$ $\rightarrow$ $\swarrow$ \\
$(-,0)|(-,+)_{|\alpha|> \alpha_c}$ &  {\Huge $\uparrow$}$\nwarrow$ $\rightarrow$ $\nearrow$ \\
$(-,0)|(-,+)_{|\alpha|= \alpha_c}$ &  {\Huge $\uparrow$} $\nwarrow$ $\rightarrow$ $0$ \\
\hline
$(-,+)|(-,0)_{|\alpha|< \alpha'_c}$ &  {\Huge $\uparrow$} $\nearrow$ $\leftarrow$ $\searrow$ \\
$(-,+)|(-,0)_{|\alpha|> \alpha'_c}$ &  {\Huge $\uparrow$}$\nearrow$ $\leftarrow$ $\nwarrow$ \\
$(-,+)|(-,0)_{|\alpha|= \alpha'_c}$ &  {\Huge $\uparrow$} $\nearrow$ $\leftarrow$ $0$ \\
\hline
$(-,+)|(-,+)_{|\alpha|< |\alpha'_c-\alpha_c|}$ &   {\Huge $\uparrow$} $\downarrow$ $\nearrow$ $\nwarrow$ $\rightarrow$ $\searrow$ $\swarrow$ \\
$(-,+)|(-,+)_{|\alpha|= \alpha'_c-\alpha_c>0}$ & {\Huge $\uparrow$} $\downarrow$  $\nwarrow$ $\rightarrow$ $\swarrow$ $0$ \\
$(-,+)|(-,+)_{|\alpha|= \alpha_c-\alpha'_c> 0}$ & {\Huge $\uparrow$} $\downarrow$ $\nearrow$ $\leftarrow$ $\searrow$ $0$ \\
$(-,+)|(-,+)_{|\alpha|= |\alpha'_c-\alpha_c|= 0}$ &   {\Huge $\uparrow$}  $\downarrow$ $0$\\
$(-,+)|(-,+)_{|\alpha'_c-\alpha_c|< |\alpha|< \alpha'_c+\alpha_c}$ & 
({\Huge $\uparrow$}) $\downarrow$ $\nearrow$ $\nwarrow$  
({\Huge $\leftarrow$}) ({\Huge $\rightarrow$}) $\searrow$ $\swarrow$ \\
$(-,+)|(-,+)_{|\alpha|= \alpha'_c+\alpha_c}$ & ({\Huge $\uparrow$}) $\nearrow$ $\nwarrow$  
({\Huge $\leftarrow$}) ({\Huge $\rightarrow$}) $0$ \\
$(-,+)|(-,+)_{|\alpha|> \alpha'_c+\alpha_c}$ &  ({\Huge $\uparrow$}) $\nearrow$ $\nwarrow$  
({\Huge $\leftarrow$}) ({\Huge $\rightarrow$})  \\
\hline
\hline
\end{tabular}
\caption{\small The feasible causal orientations of $\vec{\bm{K}}|_{x}(\vec v,\vec v)$ for the {\em contracting} cases. The 9 possibilities appearing at the right lower box of Table \ref{t3} are represented here. The critical values for the orientation parameter $\alpha$ as well as the special notation with round brackets are as in Table \ref{t5}.  \label{t10}}
\end{center}
\end{table}

\subsubsection{Stationary points}
This is probably the simplest type. Among its four possible cases, which appear at the central box of table \ref{t3}, only one has a relevant $\alpha\neq 0$: $(\pm)|(\pm)_{\alpha}$. The mean tendency for $\vec{\bm{K}}|_{x}(\vec v,\vec v)$ is to vanish, but this happens by compensation between opposite orientations rather than by vanishing $\vec{\bm{K}}|_{x}(\vec v,\vec v)$, and there are some relevant cases with no $\vec v\in T_{x}S$ such that this vanishes. The possibilities are studied in the Appendix and the results presented in Table \ref{t8}.

\subsubsection{Mixed points}
In this case the mean tendency is for $\vec{\bm{K}}|_{x}(\vec v,\vec v)$ to be spacelike. There is a correspondence between this case and the expanding one ---see the Appendix---, and the final results are summarized in Table \ref{t7}.

\subsubsection{Contracting and semi-contracting points}
These can be seen to be equivalent to the expanding and semi-expanding points with the reverse time orientation. Thus, the different possibilities are simply summarized in tables \ref{t10} and \ref{t9} respectively.

\section{The primary global classification}
\label{sec:primary}
The global cases are probably the more interesting parts of the classification. Hitherto, everything has been performed at an arbitrary but fixed point of $S$. Now the question is to see how this may vary from point to point, that is, how many different types of points exist on a particular surface. There are many feasible routes to address this problem, and the number of possibilities is enormous, as can be easily guessed from Tables \ref{t3}-\ref{t10}. Some remarkable possibilities, for instance, arise by using the horizontal rows, or the vertical columns, in Table \ref{t3} to obtain the primary classification. Even though this will not be the route used herein, there are two outstanding cases within these classifications which will be taken into account, {\em and combined with} the actual classification to be constructed in what follows. These are the following:
\begin{itemize}
\item {\em Sub-geodesic surfaces}. These are the cases contained in the 4th row, or the 4th column, of Table \ref{t3}. If every $x\in S$ is of one of the types in the 4th row (respectively the 4th column), then $S$ is called $\vec\ell$-subgeodesic (resp.\ $\vec k$-subgeodesic). For these surfaces, the affinely parametrized geodesics {\em within} $S$ ---that is to say, such that its velocity vector field satisfies $\overline\nabla_{\vec X}\, \vec X= 0$--- is a sub-geodesic \cite{Scho} with respect to $\vec \ell$ (respectively $\vec k$ ) on the manifold:
\bean
\nabla_{\vec X}\, \vec X = a\,  \vec \ell\, , \hspace{1cm} 
(a=\bm{K}_{\vec\ell}(\vec X,\vec X)),\\
\hspace{-1cm} \mbox{respectively} \hspace{1cm} \nabla_{\vec X}\, \vec X = b\,  \vec k\, , \hspace{1cm} 
(b=\bm{K}_{\vec k}(\vec X,\vec X)).
\eean
\item {\em Umbilical surfaces}. These are defined by a shape tensor which is proportional to the first fundamental form:
\be
\vec{\bm{K}}=\frac{1}{2}\vec H \, \bm{\gamma} \, .
\ee
They correspond to the surfaces such that both $\bm{K}_{\vec k}$ and $\bm{K}_{\vec\ell} $ have a {\em degenerate} type at {\em every} point. 
\end{itemize}

The route chosen here, however, is based on already existing types of surfaces (trapped, stationary, non-trapped) which are of special relevance in gravitational physics, so that they can be recovered in the new classification. The single cases will not all be named or explicitly displayed, but a {\em method} and a notation will be put forward which allows to identify and write down {\em all} possibilities.

To that end, the primary classification is based on the types presented in Table \ref{t2}. The letters 
\begin{center}
E, sE, S, M, sC, and C 
\end{center}
will be used to denote each of those cases, with the obvious correspondence. If the two 
semi-expanding cases must be considered for a single surface, or if the two null directions $\vec \ell$ and $\vec k$ must be distinguished, then sE and sE' will be used, corresponding to the $\vec\ell$-expanding and $\vec k$-expanding cases, respectively; and analogously with sC ($\vec\ell$-contracting) and sC'; and with M ($\vec\ell$-expanding) and M'. Then a given surface can be characterized by giving the pertinent list of letters ---corresponding to the type of points that are present on the surface---, in a row, with hyphen as separation. For instance
\begin{center}
C-sC-sC'-S, \hspace{5mm} E, \hspace{5mm} E-sE, \hspace{5mm} E-sE-sE', \hspace{5mm} E-S-C, \hspace{5mm} sE-M
\end{center}
et cetera. The order is not important in principle; it can be used though to reflect the dominant type of points by writing the more numerous one on the left, then the next, and so on.\footnote{This, of course, requires a well-defined measure to compare sets of points on $S$. This will usually be available. However, there are obvious cases, such as for instance surfaces where only one point, or a discrete set of points, is of a different type.}

An alternative but equivalent notation, which being more graphical is sometimes more intuitive, uses the same arrow symbols as in the previous section but now applied to the mean curvature vector field. In this case, the arrows will all be placed starting from an imaginary centre and, in case $\vec H$ vanishes somewhere, this is is indicated by placing a point at that centre. Thus, the previous displayed list becomes
$$
\begin{sideways}\begin{sideways}$\swarrow\dotsearrow$\end{sideways}\end{sideways}
\hspace{-6mm} \raisebox{2mm}{$\uparrow$} \,\,\,\, , \hspace{5mm} \downarrow \, , \hspace{5mm} \dsearrow \, , \hspace{5mm} \dsarrow \, , \hspace{5mm}
\raisebox{-5mm}{\begin{sideways}$\leftarrow\cdot\rightarrow$\end{sideways}} \, , \hspace{5mm} \begin{sideways}$\dswarrow$\end{sideways}
$$ 
and, if this is the preferred choice, then it is convenient to distinguish between M and M' by means of the opposite arrows $\rightarrow$ and $\leftarrow$, respectively, as already used in Tables \ref{t5}-\ref{t10}.

As is obvious, the number of different possibilities is already very big, of the order of a few hundreds ($\sim 2^9$). Notice, however, that there are impossible cases if the surface has enough differentiability. For instance, the cases
\begin{center}
E-C, \hspace{5mm} E-M, \hspace{5mm} M-M', \hspace{5mm} E-sE-M-C \hspace{2mm} etc. 
\end{center}
are impossible trivially, because the traces tr$\bm{K}_{\vec k}$ and tr$\bm{K}_{\vec \ell}$, which are called the {\em null expansions}, are continuous functions on $S$ and therefore cannot change sign if they do not vanish somewhere. Effectively, this implies that {\em only} some ``connected" cases are allowed: one must be able to follow an imaginary connected path when moving over Table \ref{t2} to cover all the letters appearing on the surface acronym. Besides, diagonal crossings are only allowed if they involve the type S, otherwise they are forbidden. This reduces the total number of cases substantially (to about two hundreds.)

A preliminary classification for surfaces is now given by the length $l$ of the acronyms ---its {\em number} of capital letters---, which determines the number of different types of points, according to Table \ref{t2}, contained in $S$. This produces {\em nine} classes, $l\in \{1,2,\dots ,9\}$, which will be called {\em pure}, {\em binary}, 
{\em ternary}, {\em quaternary}, and so on for $l=1,2,3,4,\dots$ respectively. Each of these 9 cases is divided into $\displaystyle{\left(\begin{array}{c} 9\\ l\end{array}\right)}$ possibilities, {\em minus} the corresponding number of impossible acronyms. All the pure, binary and ternary cases will be named and considered in what follows. This will provide us with a pattern usable to name all remaining cases, which will be considered only briefly for some relevant possibilities in the subsection \ref{higher}. Of course, there may be important cases which will thus not be explicitly considered, while probably {\em all} sort of surfaces are worth to be analyzed more deeply. However, in order to keep this paper within a reasonable length, this will have to be skipped.

\subsection{Pure surfaces}
All 9 types of pure surfaces have already been considered, and named, in the literature. They correspond to the following cases:
\begin{center}
\begin{tabular}{c|c|l}
Acronym & Symbol & Type of surface \\
\hline
E & $\downarrow$ & past-trapped\\
sE, sE'  & $\searrow$ or $\swarrow$ &marginally past-trapped \\
S & $\cdot$ & stationary or extremal (minimal/maximal) \\
M, M' & $\rightarrow$ or $\leftarrow$ & untrapped (or absolutely non-trapped) \\
sC, sC' & $\nwarrow$ or $\nearrow$ &  marginally future-trapped \\
C & $\uparrow$ & future-trapped\\
\hline
\end{tabular}
\end{center}
The only generically stable types (under {\em generic} small perturbations), are E, C, M or M'.
Each of these nine classes will be divided into the corresponding possibilities, according to Table \ref{t3}, in a secondary classification below, section \ref{sec:secondary}. These will also be refined in the final ``finer" classification of subsection \ref{sec:finer}.

\subsection{Binary surfaces}
There are $\displaystyle{\left(\begin{array}{c} 9\\ 2\end{array}\right)-20=16}$ different types of binary surfaces. A few of the binary and ternary surfaces have been considered in the literature, some others appear to have no name so far. I have named them trying to comply and be coherent with the already existing and used nomenclature. The binary surfaces are presented in the following list:
\begin{center}
\begin{tabular}{c|c|l}
Acronym & Symbol & Type of surface \\
\hline
E-sE, E-sE' & $\dsearrow$ or $\dswarrow$ & almost past-trapped\\
E-S & $\dot\downarrow$ & partly past-trapped\\ 
sE-S, sE'-S & $\dotsearrow$ or $\dotswarrow$ & partly marginally past-trapped\\
sE-M, sE'-M' & \begin{sideways}$\dswarrow$\end{sideways} or \begin{sideways}$\unwarrow$\end{sideways} & past almost untrapped (or past non-trapped)\\
M-S, M'-S & $\cdot\rightarrow$ or $\leftarrow\cdot$ & (special) partly untrapped  \\
sC'-M, sC-M' & \begin{sideways}$\dsearrow$\end{sideways} or \begin{sideways}$\unearrow$\end{sideways} & future almost untrapped  (or future non-trapped) \\
sC-S, sC'-S & \begin{sideways}\begin{sideways}{$\dotsearrow$}\end{sideways}\end{sideways} \, or \begin{sideways}{$\dotsearrow$}\end{sideways} & partly marginally future-trapped \\
C-S & \begin{sideways}\begin{sideways}$\dot\downarrow$\end{sideways}\end{sideways} & partly future-trapped\\
C-sC, C-sC' & $\unwarrow$ or  $\unearrow$ & almost future-trapped\\
\hline
\end{tabular}
\end{center}
There are no generically stable binary surfaces.
Observe that, if one takes into account the order of the letters, as explained above, then surfaces such as (say) S-E will be closer or more similar to stationary surfaces, while E-S may resemble past-trapped surfaces. This can be extended to all types. 

The partly marginally trapped cases, future and past, have often been included in the marginally trapped corresponding family, because they have a mean curvature vector which is null and proportional to one of $\vec k$ or $\vec\ell$ everywhere, but with the proportionality factor vanishing somewhere on $S$. It may be important to keep the distinction, though, especially in cases such as S-sE (ordered), which occasionally may have properties similar to stationary surfaces but not to marginally-trapped ones. 

Partly trapped surfaces have a mean curvature vector field $\vec H$ which is timelike or vanishing, keeping the causal orientation everywhere on $S$, while the partly untrapped surfaces have an $\vec H$ which is spacelike or vanishing everywhere. The use of the adverb ``partly" is restricted to these sort of cases in which the non-explicitly mentioned part is constituted by {\em stationary} points exclusively. In other words, ``partly future-trapped" (say) means a surface which is {\em partly future-trapped and partly stationary}. The partly untrapped case is dubbed as ``special'' because the generic partly untrapped surface is ternary (M-S-M', see below). The special case has semi-definite null expansions  on $S$.

The almost untrapped cases are characterized by having one definite null expansion, and the other oppositely semi-definite. Thus, they have different signs everywhere on $S$ ---considering {\em three} signs $\{+,0,-\}$---, and one of them keeps a constant non-zero sign. These cases have an $\vec H$ which is spacelike or null (with a fixed orientation) everywhere.

Almost trapped surfaces are defined by having one null expansion non-negative (respectively non-positive) and the other positive (resp.\ negative), everywhere on $S$.

As before, all these classes will be further sub-divided, according to Table \ref{t3}, in the secondary and ``finer" classifications below---section \ref{sec:secondary}.

\subsection{Ternary surfaces}
There are $\displaystyle{\left(\begin{array}{c} 9\\ 3\end{array}\right)-48=36}$ different types of ternary surfaces. A majority seem to have no name so far. All types and their used, or proposed, names are presented next.
\begin{center}
\begin{tabular}{c|c|l}
Acronym & Symbol & Type of surface \\
\hline
E-sE-sE' & $\dsarrow$ & weakly past-trapped \\
E-sE-S, E-sE'-S & $\dotdsearrow$ or $\dotdswarrow$ & feebly past-trapped\\
sE-sE'-S & $\swarrow \dotsearrow$ & null past-trapped\\
E-sE-M, E-sE'-M' & $\dsearrow$\hspace{-4mm}\raisebox{2mm}{$\rightarrow$} or \raisebox{2mm}{$\leftarrow$} \hspace{-5.5mm}$\dswarrow$  & half diverging or half past-trapped\\
E-S-sC, E-S-sC' & \raisebox{-3mm}{\begin{sideways}\raisebox{-1.5mm}{$\leftarrow$} \!\begin{sideways}{$\dotsearrow$}\end{sideways} \end{sideways}}  or 
\raisebox{-3mm}{\begin{sideways}\raisebox{3mm}{$\leftarrow$} $\dotsearrow$\end{sideways}} & causal weakly half diverging (or p-t)\\
E-S-M, E-S-M' & $\dot\downarrow$\raisebox{3mm}{$\rightarrow$} or \raisebox{3mm}{$\leftarrow$}$\dot\downarrow$ & non-null feebly half diverging (or p-t)\\
M-S-sE', M'-S-sE & $\dotswarrow$ \raisebox{3mm}{$\rightarrow$} or \raisebox{3mm}{$\leftarrow$} $\dotsearrow$ & non-timelike weakly half diverging (or p-t)\\
M-sE-S, M'-sE'-S & \begin{sideways}$\dotdswarrow$\end{sideways} or \begin{sideways}\begin{sideways}\begin{sideways}$\dotdsearrow$\end{sideways}\end{sideways}\end{sideways} & past feebly untrapped \\
M-sE-sC', M'-sE'-sC &  \raisebox{-3mm}{\begin{sideways}$\dsarrow$\end{sideways}} or \raisebox{-3mm}{\begin{sideways}$\unarrow$\end{sideways}} & weakly untrapped \\
sE-S-sC', sE'-S-sC &  \raisebox{-3mm}{\begin{sideways}$\swarrow\dotsearrow$\end{sideways}} or  \raisebox{-3mm}{\begin{sideways}\begin{sideways}\begin{sideways}$\swarrow\dotsearrow$
\end{sideways}\end{sideways}\end{sideways}}  & null untrapped\\ 
E-S-C & \raisebox{-5mm}{\begin{sideways}$\leftarrow\cdot\rightarrow$\end{sideways}} &  timelike dual \\
sE-S-sC, sE'-S-sC' & \raisebox{-2mm}{\raisebox{5.5mm}{$\nwarrow$} $\dotsearrow$} or \raisebox{-2mm}{$\dotswarrow$ \raisebox{5.5mm}{$\nearrow$}} & null dual (or marginally half trapped)\\
M-S-M' & $\leftarrow\cdot\rightarrow$ & (generic) partly untrapped \\
M-sC'-S, M'-sC-S  & \begin{sideways}$\dotdsearrow$\end{sideways} or \begin{sideways}\begin{sideways}\begin{sideways}$\dotdswarrow$\end{sideways}\end{sideways}\end{sideways}  & future feebly untrapped\\ 
M-S-sC, M'-S-sC' &  \begin{sideways}\begin{sideways}{$\dotsearrow$}\end{sideways}\end{sideways} \raisebox{-1mm}{$\rightarrow$} or \raisebox{-1.5mm}{$\leftarrow$} \!\begin{sideways}{$\dotsearrow$}\end{sideways} & non-timelike weakly half converging (or f-t)\\
C-S-M, C-S-M' & \begin{sideways}\begin{sideways}$\dot\downarrow$\end{sideways}\end{sideways}\raisebox{-1mm}{$\rightarrow$} or
\raisebox{-1mm}{$\leftarrow$}\begin{sideways}\begin{sideways}$\dot\downarrow$\end{sideways}\end{sideways} & non-null feebly half converging (or f-t)\\
C-S-sE, C-S-sE' & \raisebox{-3mm}{\begin{sideways}$\dotswarrow$ \raisebox{3mm}{$\rightarrow$}\end{sideways}} or \raisebox{-3mm}{\begin{sideways}\begin{sideways}\begin{sideways}{$\dotsearrow$}\end{sideways}\end{sideways} \raisebox{-1mm}{$\rightarrow$}\end{sideways}} & causal weakly half converging (or f-t)\\
C-sC'-M, C-sC-M' & $\unearrow$\hspace{-4mm}\raisebox{-1.8mm}{$\rightarrow$} or \raisebox{-1.8mm}{$\leftarrow$}\hspace{-4mm}$\unwarrow$ & half converging or half future-trapped\\
sC-sC'-S & \begin{sideways}\begin{sideways}$\swarrow\dotsearrow$\end{sideways}\end{sideways} & 
null future-trapped\\
C-sC-S, C-sC'-S & \begin{sideways}\begin{sideways}$\dotdsearrow$\end{sideways}\end{sideways} or  \begin{sideways}\begin{sideways}$\dotdswarrow$\end{sideways}\end{sideways} & feebly future-trapped \\
C-sC-sC' & $\unarrow$ & weakly future-trapped \\
\end{tabular}
\end{center}
The generically stable ternary surfaces are E-sE-M, E-sE'-M', C-sC'-M and C-sC-M' (though in some particular situations the perturbed $S$ may become an appropriate binary, or even a pure, surface).

The weakly trapped cases are common in the General Relativity literature and are defined by having both null expansions non-negative, or non-positive, everywhere on $S$, and non-vanishing simultaneously. Thus, $\vec H$ is non-spacelike (timelike or non-zero null) and with the same causal orientation everywhere on $S$. 

Feebly trapped surfaces are subtly different from the previous: they also have the property that both traces are semi-definite on $S$, but they {\em do} vanish simultaneously at some points of $S$. Furthermore, one of the expansions vanishes {\em only} at those intersection points. Hence, $\vec H$ is causal (timelike or null) and with the same causal orientation everywhere on $S$.

Null trapped surfaces have a consistently oriented null $\vec H$ everywhere. So, both null expansions are semi-definite on $S$ {\em and}, at each point of $S$, there is always one of the two which vanishes.

Weakly, feebly and null untrapped surfaces are analogous to the previous cases by reversing one of the signs. The three cases have both expansions {\em oppositely} semi-definite on $S$. In the first case, they never vanish simultaneously; in the second case one of the expansions vanishes at points where the other is also zero; in the third case, at least one of the expansions vanishes at each point of $S$. Alternatively, the first case has a mean curvature vector which is spacelike or non-vanishing null everywhere; the second case has $\vec H$ spacelike or null everywhere. When it is null, its causal orientation is always the same, and thus one can subdivide them in past and future; and the third case has $\vec H$ null everywhere (with changing causal orientation).

The ``half" cases are also common in the GR literature, but they are usually termed as ``inner'' or ``outer'' trapped. This is either because there is a well-defined notion of outer direction, or because one is loosely speaking and determines by decree that the appropriate ``half'' direction will be called inner (or outer). If any reader is familiar with this outer/inner nomenclature, then in what follows he/she may mentally replace every ``half" by ``outer" or ``inner", and at the same time every ``diverging" by ``past-trapped" and every ``converging" by ``future-trapped". Mathematically, the half cases are characterized by having just one of the expansions definite (or semi-definite in the weak/feeble cases). They can be treated as a single family {\em if} one may only care about one of the normal null directions, due to physical or whatever else reasons. This, however, is not a very refined approach, as there are quite diverse possibilities depending not only on the sign of the other expansion on $S$, but also on their  {\em mutual} behaviour. There are 16 ``half" cases in total, but they can be individually defined in an invariant way. 

To start with, all of these cases except four are such that $\vec H$ is non-zero spacelike at a non-empty subset of $S$. The four exceptions are given by E-S-sC', E-S-sC, C-S-sE and C-S-sE'. These cases have a mean curvature vector field which is {\em causal} everywhere. However, its causal orientation changes on $S$, and for the first two cases it is past-pointing whenever $\vec H$ is timelike, while it is future-pointing whenever $\vec H$ is null. Reversely for the other two cases. Therefore, these four cases may be called ``causal weakly half" diverging or converging according to the orientation of the timelike $\vec H$'s.

There are also four `half' cases with one definite expansion: E-sE-M, E-sE-M', C-sC'-M and C-sC-M'. These are simply called half diverging (for the first two) or half converging (the second pair). The mean curvature vector is causal (with a unique causal orientation) or non-vanishing spacelike everywhere on $S$.

Four other `half' cases (E-S-M, E-S-M', C-S-M, and C-S-M') have a mean curvature vector which is timelike (with a consistent orientation) or spacelike everywhere. These are feebly (rather than weakly) half diverging or converging, because the two expansions vanish exactly at the same subset of $S$. They will be termed ``non-null feebly half" because $\vec H$ is non-null everywhere.

The remaining four `half' cases (M-S-sE', M'-S-sE, M-S-sC, and M'-S-sC') have an $\vec H$ which is spacelike or null (with only one causal orientation) everywhere. Following the same rules as before they will be labelled ``non-timelike weakly half" as their $\vec H$ is non-timelike on all points of $S$.

The partly untrapped surfaces have an $\vec H$ which is spacelike or vanishing everywhere, and they are partly untrapped and partly stationary. M-S-M' surfaces are the generically partly untrapped ones (there are special ones: M-S and M'-S). In this generic case, both null expansions have opposite signs at every point of $S$ (or they vanish simultaneously), and also all signs are realized for both of them.

The null dual surfaces have a null mean curvature vector which points consistently along one of $\vec k$ or $\vec \ell$ everywhere, but its causal orientation changes on $S$. Equivalently, one of the null expansions is identically zero on $S$, the other takes all possible signs. Therefore, they may also be called marginally half trapped.

Finally, probably the most exotic ternary case is E-S-C. The mean curvature vector is timelike (or zero) everywhere on $S$, but both causal orientations, future and past, are realized at different subsets of $S$. Due to this distinctive property, they will be called ``timelike dual" surfaces. Both expansions have the same sign everyhwere on $S$, and they take all three signs. 

As before, every ternary case can be subdivided according to the cases in Table \ref{t3}.

\subsection{Some quaternary and higher surfaces}
\label{higher}
For quaternary and higher surfaces the basic ideas are the same as for the three cases studied, and there are no significant new behaviours: just combinations of the previous types. Thus, once the pattern of names and properties have been explicitly shown, every case can be dealt with easily. Only a few outstanding cases will be considered explicitly here:
\begin{itemize}
\item Nearly trapped surfaces (\begin{sideways}\begin{sideways}$\swarrow\dotsearrow$
\end{sideways}\end{sideways}\hspace{-7.8mm} \raisebox{2mm}{$\uparrow$}\hspace{2mm} , $\swarrow\! \dotsearrow$
\hspace{-9mm} \raisebox{-1mm}{$\downarrow$}\hspace{7mm}): these are defined \cite{MS} by having a mean curvature vector which is causal and with a fixed causal orientation all over $S$. They are E-sE-sE'-S (nearly future trapped) and C-sC-sC'-C (nearly past trapped). From the null-expansion point of view, both of them are semi-definite (with the same sign) everywhere on $S$.
\item Nearly untrapped surfaces ( 
\raisebox{-3mm}{\begin{sideways}$\swarrow\dotsearrow$
\end{sideways}\hspace{-5mm} \raisebox{4mm}{$\rightarrow$}},  
\raisebox{0.5mm}{$\leftarrow$}\hspace{-3mm}\raisebox{-3mm}{\begin{sideways}\begin{sideways}\begin{sideways}
$\swarrow\dotsearrow$\end{sideways}\end{sideways}\end{sideways}} ): analogously one can consider the cases with both expansions semi-definite, but with opposite signs, on $S$. These have a mean curvature vector which is spacelike or null (with both orientations) everywhere. These types are M-sE-sC'-S and M'-sE'-sC-S.
\item 7-ary surfaces: There is a particularly interesting case here: E-sE-sE'-S-sC-sC'-C. In other words, the only missing letters are M and M'. Its graphical symbol is
$$
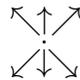
\stackrel{\begin{sideways}\begin{sideways}$\swarrow\dotsearrow$
\end{sideways}\end{sideways}\hspace{-6mm} \raisebox{2mm}{$\uparrow$}}
{\hspace{3.5mm}\swarrow\!\downarrow\! \searrow} 
$$
These surfaces, {\em as well as all their subcases} obtained by eliminating some of the letters in the acronym (or some of the arrows and/or the point in the symbol), are defined by having a non-spacelike mean curvature vector everywhere on $S$. Therefore, $g(\vec H,\vec H)$ is non-positive all over $S$. This may have relevance in some applications (section \ref{sec:apps}). Thus, the generic family with this property, as well as all its sub-cases ---which include in particular all the (null, feebly, weakly, nearly, almost, partly) trapped or marginally trapped surfaces, and the dual surfaces too, among many others--- will be given the common graphical name of $\varhexstar$-{\em surfaces}.
\item 8-ary surfaces: these are characterized by the missing letter in the acronym, rather than by the present ones. There are, thus, 9 types of these surfaces, and another possible notation would be ``$\not\mbox{E}$", or also ``not-E", say. 
\item Generic surface: E-sE-M-sE'-S-sC'-M'-sC-C. It contains all type of points. Its symbol is
$$
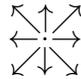
\raisebox{3mm}{$\leftarrow$} \hspace{-8.5mm} \stackrel{\begin{sideways}\begin{sideways}$\swarrow\dotsearrow$
\end{sideways}\end{sideways}\hspace{-6mm} \raisebox{2mm}{$\uparrow$}}
{\hspace{3.5mm}\swarrow\!\downarrow\! \searrow} \hspace{-5mm}\raisebox{3mm}{$\rightarrow$}
$$
and it is obviously generically stable.
\end{itemize}

\section{The secondary global classifications}
\label{sec:secondary}
Each of the cases considered in the previous section can be sub-classified according to the different types of points appearing in Table \ref{t3}. Thus, each letter (E, C, S, et cetera) can be refined by considering the types of points in each of the corresponding 9 boxes delimited by double lines in Table \ref{t3}. This leads to the secondary classification of surfaces. As before, some of the cases are historically well-known (e.g. umbilical surfaces, or totally geodesic surfaces) but the majority of cases had not been explicitly named before.

For the secondary classification it is probably useless to consider all possibilities, that is, surfaces with one type of points, with two, and so on. Rather, what seems to be logical and adapted to the primary classification is to sort out the surfaces according to a hierarchy of the types of points which are feasible for each of the letters in the acronym. These correspond to different entries of Table \ref{t3}, and the hierarchy to be used is the one shown on each of the Tables \ref{t5}-\ref{t10} ignoring $\alpha$, so that the highest level corresponds to the upper case in each table, and the increasing direction is upwards on all tables. Thus, there will only be 6 cases for either E, M, M', C, sE, sE', sC, or sC' (for the former four the 9 initial cases are effectively reduced to 6 due to the symmetry between $\vec\ell$ and $\vec k$); and 4 cases for stationary points of type S (in this case it is actually convenient to consider the 4 entries in the middle box of Table \ref{t3} separately). Then, the type of surface in the secondary global classification will correspond, for each letter in the acronym, to the minimum type of point (for that letter) contained in the given surface $S$. 

The following explicit division, which comprises the pure surfaces and is the basis for the rest of the secondary global classification, will serve as clarifying illustrative examples.

\subsection{Pure surfaces}
Consider the pure surfaces except the stationary case S. Each of these surfaces can be sub-classified into six classes according to the less specialized type of point contained in the surface. The adverbs {\em merely, very lightly, lightly, strongly, very strongly} and {\em totally} will be used before the corresponding name to denote the type of surface. Thus, for instance, the past-trapped surfaces E can be divided as follows:
\begin{enumerate}
\item If $S$ contains points of type $(+,-)|(+,-)$ then $S$ is called {\em merely} past-trapped.
\item If $S$ does not contain points of type $(+,-)|(+,-)$, but does have points of type $(+,-)|(+,0)$ or $(+,0)|(+,-)$, then $S$ is called {\em very lightly} past-trapped.
\item If $S$ does not contain points of type $(+,-)|(+,0)$ or $(+,0)|(+,-)$ or lower, but has points of type $(+,-)|(+,+)$ or $(+,+)|(+,-)$, then $S$ is said to be {\em lightly} past-trapped.
\item If $S$ contains points only of type $(+,0)|(+,0)$ or higher then $S$ is called {\em strongly}  past-trapped (also, ``future semi-convex''.)
\item If $S$ solely contains points of type $(+,+)|(+,0)$, $(+,0)|(+,+)$ and $(+,+)|(+,+)$ then $S$ is called {\em very strongly} past-trapped (also, ``future convex''.)
\item If $S$ consists of points of type $(+,+)|(+,+)$ exclusively, then $S$ is called {\em totally} past-trapped (also, ``future strictly convex''.) They contain the {\em past-trapped umbilical} surfaces, which have type $(+)|(+)$ all over $S$.
\end{enumerate}
and analogously for future-trapped (C), marginally trapped (sE, sE', sC and sC'), and untrapped (M and M') surfaces. 


The remaining case S, the stationary surfaces, is somewhat special and therefore we use specific different names:
\begin{enumerate}
\item If $S$ contains points of type $(\pm)|(\pm)$ then $S$ is called {\em simply} stationary.
\item If $S$ is not simply stationary, and has points of type $(\pm)|(0)$ {\em and} $(0)|(\pm)$, then $S$ is said to be {\em mildly} stationary or {\em compound}.
\item If $S$ contains points of type $(\pm)|(0)$ and $(0|0)$ exclusively (respectively of type $(0)|(\pm)$ and $(0|0)$ exclusively) then $S$ is called {\em $\vec k$-subgeodesic}  (respectively, {\em $\vec\ell$-subgeodesic}) and stationary. Both possibilities will also be termed as {\em highly} stationary.
\item If all points in $S$ are of type $(0)|(0)$, then $S$ is called {\em totally geodesic} ---also {\em fully} stationary.
\end{enumerate}
The names ``fully" and ``highly" stationary are included here for convenience in order to be able to produce names for binary and higher-order surfaces, but they are not really necessary ---not even convenient in the former case--- for the pure surfaces, as their first names are more informative and in one case traditionally used.


A symbolic notation is easily devised at this stage for these pure surfaces at the secondary level. It is enough to add, as a subscript to the symbol in the primary classification, the ``minimum" type of point existing in the surface. Schematically
$$
\mbox{(primary symbol for $S$)}_{\mbox{\scriptsize less-specialized type of point in $S$}}\, \, .
$$
For instance,
$$
\uparrow_{(-,0)|(-,0)}\, ; \hspace{2mm} \nearrow_{(\pm)|(-,-)}\, ; \hspace{2mm} \leftarrow_{(-,+)|(+,-)}\, ; \hspace{2mm} \cdot_{(\pm)|(\pm)}\, ; \hspace{2mm} \downarrow_{(+,+)|(+,+)}
$$
denote the surfaces which are strongly future-trapped, lightly marginally future-trapped, merely untrapped, simply stationary, and totally past trapped, respectively.

Some particular cases are noticeable and worth mentioning. These are given in the following list with their symbols:
\begin{center}
\begin{tabular}{c|l}
Symbol & Type of surface \\
\hline
$\downarrow_{(+)|(+)}$ & past-trapped umbilical\\
$\searrow_{(+,-)|(0)}$ & $\vec k$-subgeodesic marginally past-trapped\\
$\searrow_{(+)|(0)}$ & $\vec k$-subgeodesic past umbilical\\
$\swarrow_{(0)|(+,-)}$ & $\vec \ell$-subgeodesic marginally past-trapped\\
$\swarrow_{(0)|(+)}$ & $\vec \ell$-subgeodesic past umbilical\\
$ \cdot_{(\pm)|(0)}$ & $\vec k$-subgeodesic stationary\\
$ \cdot_{(0)|(\pm)}$ & $\vec \ell$-subgeodesic stationary\\
$\cdot_{(0)|(0)}$ & totally geodesic \\
$\rightarrow_{(+)|(-)}$, $\leftarrow_{(-)|(+)}$ & untrapped umbilical \\
$\nearrow_{(0)|(-)}$ & $\vec \ell$-subgeodesic future umbilical\\
$\nearrow_{(0)|(-,+)}$ & $\vec \ell$-subgeodesic marginally future-trapped\\
$\nwarrow_{(-)|(0)}$ & $\vec k$-subgeodesic future umbilical\\
$\nwarrow_{(-,+)|(0)}$ & $\vec k$-subgeodesic marginally future-trapped\\
$\uparrow_{(-)|(-)}$ & future-trapped umbilical\\
\hline
\end{tabular}
\end{center}

Even though the above seems a reasonable classification, sometimes it is useful to know the particular types of points that are actually present on a given surface, and not only the less specialized one. To that end, a full list of these types of points must be provided. This is achieved by adding this list between braces as the susbscript as follows
$$
\mbox{(primary symbol for $S$)}_{\{\mbox{\scriptsize list of types of points in $S$}\}}\, \, .
$$
Therefore, one can consider (say) the following symbols as equivalent
$$
\nearrow_{(\pm)|(-,-)} \hspace{3mm} \equiv \hspace{3mm} 
\nearrow_{\{(\pm)|(-,-),(0)|(-,+),(0)|(-,0),(0)|(-,-)\}}
$$
but there are lightly marginally future-trapped surfaces of more special kind such as
$$
\nearrow_{\{(0)|(-,0),(0)|(-,-)\}} \hspace{1cm} \mbox{or even (say)}  \hspace{5mm}
\nearrow_{\{(0)|(-,0)\}}\, .
$$

\subsection{Binary and higher surfaces}
These can be arranged according to the previous classification for pure surfaces, by using another hierarchy for the letters used in the acronyms (or the corresponding arrow symbols). The new hierarchy is given by
$$
\{\mbox{S}\} < \{\mbox{sE, sE', sC, sC'}\} < \{\mbox{E, C, M, M'}\} \, .
$$
Hence, it is sometimes enough to denote the surfaces by its symbol and, as a subscript, the type of less-specialized point in the ``lower letter". This is well defined whenever
\begin{itemize}
\item there is only one causal orientation (future or past) for $\vec H$, so that its symbol has no arrows pointing downwards, or no arrows pointing upwards. These are the surfaces E-sE-sE'-S-M-M'  or C-sC-sC'-S-M-M', and all their sub-cases by deleting letters on these acronyms but keeping at least one causally oriented letter. In these cases one has to add the less-specialized type of point of type S if they are present; if not, the lower type of point for sE or sE' for the past case, and analogously for the future cases. 
\item $\vec H$ is not causally oriented, that is, it is spacelike all over $S$. These are the partly untrapped surfaces and its subcases, including the untrapped $S$. 
\end{itemize}
For all the remaining cases, which are characterized by having an $\vec H$ realizing both causal orientations (future and past) on $S$, the best idea is to add {\em two} labels to the primary symbol (as a subscript and a superscript, say), one corresponding to the lower type of ``past" point, the other to the lower type of ``future" point. The lower type of point for the stationary part of the surface could also be added, then having three labels.

Concerning the names, one can add the adverb(s) corresponding to the lower type of point as used for pure surfaces before. Admittedly, this may become cumbersome and messy, but there is no obvious simplification if one wishes a complete and detailed classification.

Several examples are in order. One can write, for instance,
$$
\dsarrow_{(\pm)|(+,0)}\, ;  \hspace{7mm} \raisebox{-5mm}{\begin{sideways}$\leftarrow\cdot\rightarrow$\end{sideways}}_{(+,-)|(+,0)}^{(-,-)|(-,-)}\, ;  \hspace{7mm} \begin{sideways}\begin{sideways}$\swarrow\dotsearrow$
\end{sideways}\end{sideways}\hspace{-6mm} \raisebox{2mm}{$\uparrow$}_{(\pm)|(\pm)}
$$
and the first one might be called very-lightly weakly past-trapped, while the second may be termed ``very-lightly-past totally-future timelike dual"; finally the third could be called ``simply nearly future-trapped''. Observe that the generic $\varhexstar$-surfaces are those with type
$$
\stackrel{\begin{sideways}\begin{sideways}$\swarrow\dotsearrow$
\end{sideways}\end{sideways}\hspace{-6mm} \raisebox{2mm}{$\uparrow$}}
{\hspace{3.5mm}\swarrow\!\downarrow\! \searrow}_{(\pm)|(\pm)} \, .
$$

Note finally that a complete classification and a detailed symbol can be achieved by simply adding either (i) the lower type of point or (ii) the full list of types of points between braces, for {\em each} letter in the acronym, that is to say, by labelling each arrow with its corresponding less-specialized point or list of points. 
(There may arise pictorial difficulties by adding the corresponding labels ... !) Some examples would be
$$
{}_{(+,+)|(+,0)}\!\dsearrow\hspace{-5mm}\raisebox{2mm}{$\rightarrow$}_{(+,-)|(\pm)}^{(+,-)|(-,+)}\, \, , \hspace{15mm}
{}_{(+,+)|(+,0)}\!\dsearrow\hspace{-5mm}\raisebox{2mm}{$\rightarrow$}_{\{(+,+)|(0)\}}^{\{(+,+)|(-,0),(+,+)|(-,-)\}}
$$
where it should be clear that the second shown symbol corresponds to a particular case of the first one. 
Using this notation the timelike dual umbilical surfaces are given by
$$
\raisebox{-5mm}{\begin{sideways}$\leftarrow\cdot\rightarrow$\end{sideways}}_{(+)|(+)}^{(-)|(-)}
\hspace{-1.1cm}{}_{(0)|(0)}
$$
while the general sub-geodesic surfaces belong to the null dual surfaces and are given by
$$
\raisebox{2mm}{$^{(-,+)|(0)}$}\raisebox{-2mm}{\raisebox{5.5mm}{$\nwarrow$} $\dotsearrow$}_{(+,-)|(0)} \hspace{-16mm}\raisebox{1mm}{${}_{(\pm)|(0)}$}\hspace{7mm}, \hspace{5mm}
\raisebox{-2mm}{$_{(0)|(+,-)}$}\raisebox{-2mm}{$\dotswarrow$ \raisebox{5.5mm}{$\nearrow$}}^{(0)|(-,+)} \hspace{-17mm}{}_{(0)|(\pm)}
$$
which can in fact be simplified to 
$$
\raisebox{-2mm}{\raisebox{5.5mm}{$\nwarrow$} $\dotsearrow$}_{|(0)} \,\,\, \, ,  \,\,\, \,
\raisebox{-2mm}{$\dotswarrow$ \raisebox{5.5mm}{$\nearrow$}}^{(0)|}
$$
by just omitting the symbols corresponding to the non-vanishing  null second fundamental form. This type of simplifications in the notation can be used at discretion.

\subsection{Finer classification}
\label{sec:finer}
Finally, the previous classifications can be refined by adding the mutual orientation of $\bm{K}_{\vec\ell}$ and $\bm{K}_{\vec k}$ over the surface $S$. This is ruled by the angle $\alpha$ introduced in subsection \ref{relative}. Obviously, $\alpha$ is a continuous function on any differentiable surface $S$. Nonetheless, its value may well change from point to point. 

Using the same methodology as before, one can then classify each surface according to the allowed values of $\alpha$ for each particular type of point in Table \ref{t3}. This will usually amount to adding the {\em range} of allowed values of $\alpha$ for each of the arrows (and sometimes for each of the types of points within one arrow) appearing on the symbol of $S$. 
The interval of allowed values of $\alpha$ can be closed, open, left-open or right-open and should be written accordingly. 

At every point where either $\bm{K}_{\vec\ell}$ and $\bm{K}_{\vec k}$ belongs to one of the degenerate cases, $\alpha$ is taken to be vanishing.
The cases with $\alpha =0$ and $\alpha=\pi/2$ everywhere on $S$ will be called {\em globally congruent} and {\em globally orthogonal}, respectively.

\section{Discussion and lines of research}
\label{sec:apps}
The purpose of this paper is just to present the extrinsic classification of surfaces in spacetimes. Nevertheless, the immediate question arises of whether or not this classification can be useful in some investigations, interesting concerning applications, or relevant in any way. This section is devoted to address this question in a very succinct manner. The conclusion is that the classification seems to be helpful and may have various physical and mathematical applications. There are also open doors for generalizations, specially to the case of higher dimensions, see subsection \ref{subsec:higher}.

To start with, I would like to very briefly sketch an example showing that, at least in the primary classification, most of the cases can actually happen. Take the manifold $\varietat = \mathbb{R}^2 \times S^1 \times S^1$ with metric
$$
g=\bm{\eta}_{2}+(b+a\cos\phi)^2 d\varphi\otimes d\varphi + a^2 d\phi\otimes d\phi
$$
where $\bm{\eta}_{2}=-(du\otimes dv+dv\otimes du)$ is the metric of flat 2-dimensional spacetime, $\phi$ and $\varphi$ are standard coordinates on the torus $S^1 \times S^1$, and $b>a>0$ are smooth functions not depending on $\phi$. For each pair of constants $u_{0},v_{0}$, the surfaces $u=u_{0}$, $v=v_{0}$ are tori whose first fundamental form is the metric of a torus with major radius $b(u_{0},v_{0},\varphi)$ and minor radius ---the radius of the sections--- $a(u_{0},v_{0},\varphi)$, both depending on the azimuth and therefore non-constant in general. Using for instance the simple formulas presented in \cite{S0}, the mean curvature one-form $\bm{H}\equiv g(\vec H, \cdot)$ is given by
$$
\bm{H}=\left.\frac{\partial U}{\partial u}\right|_{u_{0},v_{0}}du+
\left.\frac{\partial U}{\partial v}\right|_{u_{0},v_{0}}dv, \hspace{2cm} U\equiv \log [a(b+a\cos\phi)]\, .
$$
Thus, by choosing appropriately the values of $a,b$ and its first derivatives on the surface one can produce mean curvature vector fields with varying orientations of any desired type. In particular, one can construct very easily generic 9-ary surfaces.

Concerning open lines of research, an obvious one is the possible relation to other classifications such as those based on eigenvalues of the Laplacian on $S$ \cite{Chen,Chen1,Chen2}, or on differential properties of the mean curvature vector $\vec H$ on $S$, etc. 

Another almost immediate question concerns the actual existence (or absence) of a chosen type of surface in a given background space-time $\espaitemps$. Thus, for instance, one of the most important problems in gravitational physics concerns the appearance of (marginally, weakly, feebly, nearly) trapped surfaces, specially if they are {\em closed}---meaning compact without boundary---; see subsection \ref{subsec:trapped}. 
In addition to that question of (non-) existence of a particular type of surface, a possible virtue of the classification is the refinement  provided for standard classical surfaces, such as trapped, umbilical or stationary surfaces. For instance, considering the stationary surfaces (whose generic type ---called ``simply stationary'' surfaces--- is denoted by $\cdot_{(\pm)|(\pm)}$), apart from the totally geodesic case $\cdot_{(0)|(0)}$ there arise two other types of surfaces, the $\vec k$- and 
$\vec\ell$-subgeodesic stationary surfaces ($\cdot_{(\pm)|(0)}$ and $\cdot_{(0)|(\pm)}$, respectively). These surfaces, and in particular their subcases $\cdot_{\{(\pm)|(0)\}}$ and $\cdot_{\{(0)|(\pm)\}}$, seem to have not been studied as such in the literature. One can certainly find properties for them: an example would be the result in \cite{AR,AR1} that surfaces of type $\cdot_{(\pm)|(\pm)}$, $\cdot_{(\pm)|(0)}$ or $\cdot_{(0)|(\pm)}$ have a definite upper bound for its Gaussian curvature in the de Sitter space-time: above that bound all stationary surfaces must necessarily be totally geodesic.

The standard tools to obtain these (non-)existence results are the Gauss-Bonnet formula, and the Gauss, Codazzi, and Ricci equations. Take for instance the Ricci equation, which can be written as
\be
d\bm{\theta}= \mbox{Riem}(\vec\ell,\vec k,\cdot ,\cdot)^{T} -\left[\bm{K}_{\vec k},\bm{K}_{\vec\ell}\right] \label{ricci}
\ee
where I use the notation introduced in subsection \ref{relative}, Riem  is the Riemann tensor of the spacetime, and $\bm{\theta}$ is the one-form on $S$ defined by
$$
\bm{\theta}(\vec X)=g(\vec\ell,\nabla_{\vec X}\vec k) \hspace{1cm} \forall \vec X\in \mathfrak{X}(S)\, .
$$
Note that $\bm{\theta}$ is not invariant under (\ref{free})
$$
\bm{\theta}'=\bm{\theta}+d\log \sigma^2
$$
and therefore it is defined up to the addition of an exact differential. From (\ref{ricci}) follows that the parameter $\alpha$ introduced in \ref{relative}, which rules the relative orientation of the two null second fundamental forms, is a function on $S$ governed by $d\bm{\theta}$ and the appropriate components of the curvature of the spacetime. 

Suppose, for instance, that the $\espaitemps$ has constant curvature. Then the Riem term in (\ref{ricci}) vanishes identically and one deduces the following interesting result
$$
d\bm{\theta}=-\left[\bm{K}_{\vec k},\bm{K}_{\vec\ell}\right]
$$
so that the two null second fundamental forms commute if and only if $\bm{\theta}$ is closed. In particular, in constant-curvature spacetimes, the congruent and orthogonal cases ($\alpha =0,\pi /2$), as well as the $\vec k$- or $\vec\ell$-umbilical cases ---including the umbilical surfaces--- must necessarily have a closed $\bm{\theta}$, so that it can be made to vanish locally.

Consider now the Gauss equation, which can be written as
\be
2K_{S}=R+4\, \mbox{Ric}(\vec k, \vec\ell\, )-2\, \mbox{Riem}(\vec k,\vec\ell,\vec k,\vec\ell\, )+g(\vec H,\vec H\, ) -2\,  \mbox{tr}\, (\bm{K}_{\vec k}\bm{K}_{\vec \ell}) \label{gauss}
\ee
where $K_{S}$ is the Gaussian curvature of $S$, and $R$ and Ric are the scalar curvature and Ricci tensor, respectively. By using the notation introduced in previous sections, this can be rewritten as
\be
K_{S}=\frac{R}{2}+2\, \mbox{Ric}(\vec k, \vec\ell\, )-\mbox{Riem}(\vec k,\vec\ell,\vec k,\vec\ell\, )-\left[\left(\lambda_{1}\mu_{1}+\lambda_{2}\mu_{2}\right)\sin^2\alpha+
\left(\lambda_{1}\mu_{2}+\lambda_{2}\mu_{1}\right)\cos^2\alpha\right]\label{gauss2}
\ee
from where it is relatively simple to deduce many direct consequences for the different types in the classification if one imposes some restrictions on the curvature of the space-time. Observe, for instance, that the term in square brackets is positive at any point of type $(+,+)|(+,+)$ or $(+,+)|(+,0)$, and so on. As a trivial example, take the sub-geodesic surfaces, that is
$$
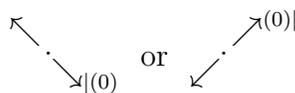
\raisebox{-2mm}{\raisebox{5.5mm}{$\nwarrow$} $\dotsearrow$}_{|(0)} \,\,\, \, \mbox{or}  \,\,\, \,
\raisebox{-2mm}{$\dotswarrow$ \raisebox{5.5mm}{$\nearrow$}}^{(0)|}
$$
then the term in square brackets in (\ref{gauss2}) vanishes on $S$. Therefore, in Minkowski flat spacetime  (Riem = 0) all sub-geodesic surfaces must be locally flat. Similarly, consider the case of stationary surfaces ($\vec H =\vec 0$) and flat space-time. The previous formula immediately implies that
$$
K_{S}=2\lambda_{1}\mu_{1}\cos 2\alpha
$$
which is positive, zero or negative for $|\alpha|$ less, equal or greater than $\pi/4$, respectively. As a matter of fact, this simple result can be strengthened and the cases with $|\alpha|<\pi /4$ everywhere on $S$ are forbidden, as follows from \cite{AP1}. Another immediate consequence is that all non-flat stationary surfaces in a flat spacetime must be {\em simply} stationary, that is, they must contain points of type $(\pm)|(\pm)$ ---and, actually, points of type $(\pm)|(\pm)_{|\alpha|>\pi/4}$.

Combining (\ref{gauss}) with the Gauss-Bonnet theorem, which reads for closed surfaces
\be
\int_{S} K_{S} =2\pi \chi(S) =4\pi (1- g_{S})\label{GB}
\ee
where $\chi(S)$ is the Euler characteristic of $S$ and $g_{S}$ its genus, one can also derive many strong results for particular types of surfaces in the classification, specially concerning the topology of closed $S$. Conversely, by assuming spherical or toroidal topology ($g_{S}=0,1$), restrictions on the type of surface certainly arise. For instance, there are no sub-geodesics spheres or tori in anti de Sitter spacetime.
In summary, the full implications of the Gauss-Bonnet and Gauss-Codazzi-Ricci formulae on the classification, and vice versa, are worth exploring.

To cite another feasible application of the classification, let us mention the question of the deformation of surfaces, and the corresponding variation of the area functional. As an example, consider the problem treated in \cite{AP}: the deformation of stationary surfaces but keeping a null $\vec H$. Using the notation introduced in previous sections, these are the deformations of stationary surfaces into a particular class of $\varhexstar$-surfaces ---for more on $\varhexstar$-surfaces, see subsection \ref{subsec:star}---, namely:
$$
\cdot_{(\pm)|(\pm)} \hspace{1cm} \stackrel{\mbox{deformation}}{\leadsto\, \leadsto\, \leadsto\, \leadsto} \hspace{1cm} 
\stackrel{\begin{sideways}\begin{sideways}$\swarrow\dotsearrow$
\end{sideways}\end{sideways}}
{\swarrow \searrow}
$$
The result in \cite{AP} is that, {\em for this particular kind of deformations}, the stationary surfaces are always locally minimizing for the area functional if the null convergence condition holds (that is, Ric($\vec N,\vec N)\geq 0$ for all null $\vec N$, \cite{HE,S}.) However, the graphical version of the classification clearly shows the strong restrictions that have been assumed to obtain this result: generic deformations will always lead to generic surfaces (that is, the 9-ary surfaces, E-sE-M-sE'-S-sC'-M'-sC-C), and the variations of the area functional should also be addressed in this case.

As a final simple example, take the most general spherically symmetric space-time. The metric can always be written as
$$
g=e^{2f}\bm{\eta}_{2}+r^2 d\Omega_{2}
$$
where is the flat 2-dimensional metric as above, $d\Omega_{2}$ is the standard metric on the round 2-sphere, and $f,r$ are smooth functions depending on $u$ and $v$ exclusively. Consider an arbitrary 2-sphere adapted to the symmetry of the space-time: $u = U$, $v = V$ for some constants $U,V$. An elementary calculation shows that
$$
\vec{\bm{K}}=\frac{1}{2}\vec H \, \bm{\gamma} = \frac{1}{2}\vec H \, r^2(U,V) d\Omega_{2}
$$
so that all these 2-spheres are umbilical, as was to be expected. Furthermore, $\vec H$ is constant on each of these 2-spheres, so that the only possible types for these preferred 2-spheres are
$$
\downarrow_{(+)|(+)}\,\,\, ,
\searrow_{(+)|(0)}\,\,\, ,
\swarrow_{(0)|(+)}\,\,\, ,
\cdot_{(0)|(0)}\,\,\, ,
\rightarrow_{(+)|(-)}\,\,\, , 
\leftarrow_{(-)|(+)}\,\,\, ,
\nearrow_{(0)|(-)}\,\,\, ,
\nwarrow_{(-)|(0)}\,\,\, ,
\uparrow_{(-)|(-)}\, .
$$
Specializing (\ref{gauss2}) to this case one also gets
$$
\frac{1}{r^2(U,V)}=\left.\left(\frac{R}{2}+2\, \mbox{Ric}(\vec k, \vec\ell\, )-\mbox{Riem}(\vec k,\vec\ell,\vec k,\vec\ell\, )\right)\right|_{S}-2\lambda_{1}\mu_{1}
$$
so that in flat Minkowski or anti de Sitter spacetimes one must have $\lambda_{1}\mu_{1}<0$: in these two spacetimes, only untrapped 2-spheres are permitted. Graphically, the only surviving possibilities are
$$
\rightarrow_{(+)|(-)}\,\,\, , \leftarrow_{(-)|(+)}\,\,\, .
$$

These are just randomly chosen, very simple, examples.

\subsection{Trapped surfaces and their relatives}
\label{subsec:trapped}
As is well known, one of the most important concepts in gravitational theories is that of a {\em closed trapped surface} introduced by Penrose in 1965 \cite{P}. It arises in the development of the singularity theorems \cite{HP,HE,S}, in the study of gravitational collapse, formation of black holes and the cosmic censorship conjecture \cite{MTW,Wald1} and the related Penrose inequality \cite{BC}, and in the recent and very interesting developing subject of marginally trapped tubes and dynamical or trapping horizons \cite{I,Hay,KH,AK}.

Apart from the traditional or known concepts of trapped, weakly trapped, nearly trapped or marginally trapped surfaces, the classification presented in this paper also shows, on the one hand, that there are other important types of related surfaces, such as the almost, partly, or feebly trapped surfaces; and on the other hand, perhaps more importantly, that each of these concepts ---including the traditional ones--- can be refined into many different sub-cases, such as the merely, very lightly, lightly, strongly, very strongly, and totally (marginally) trapped surfaces. 

In my opinion, these refinements may be relevant for the problems of gravitational collapse and dynamical or trapping horizons. For instance, I believe that the concepts of  closed very strongly  trapped surfaces, and closed totally trapped surfaces, can open new lines of research concerning the so-called ``hoop conjecture'', see e.g. \cite{T,MTW,Wald1}. Furthermore, one can ask for example if there are any restrictions on the type of closed future-trapped surfaces that form in gravitational collapse of {\em realistic} stars or galaxies, or in the closed past-trapped surfaces that are usually assumed to exist in realistic cosmological models. Similarly, there is the issue of whether or not one can improve, or strengthen, the conclusions of the singularity theorems by not merely assuming the existence of a closed trapped surface, but that of a particular type within the fauna mentioned above.

Turning back to the presently important subject of dynamical or trapping horizons, let us recall that these are essentially hypersurfaces foliated by marginally future-trapped closed surfaces (usually of spherical topology). The obvious question arises of whether or not all type of marginally future-trapped closed surfaces are allowed in generic horizons; and also how the type of surface can change along the horizon in realistic situations. If our physical intuition is correct, in most realistic and asymptotically flat cases \cite{HE} the trapping horizons will tend to be tangent  at infinity to the actual {\em event} horizon, which is a null hypersurface by definition. Therefore, it might be plausible that, in such physical systems, the trapping horizons are ruled by some sort of {\em peeling behaviour} towards infinity.

Concerning the actual existence of a concrete type of trapped surfaces, the usual techniques already mentioned can be used to prove or disprove, or even classify, the selected type of surface. For instance, one knows that there cannot be nearly trapped closed surfaces in space-times with a timelike Killing vector field \cite{MS}. This means that closed surfaces of type
$$
\begin{sideways}\begin{sideways}$\swarrow\dotsearrow$
\end{sideways}\end{sideways}\hspace{-6mm} \raisebox{2mm}{$\uparrow$}\hspace{2mm} 
$$
(or its time reversal), and its sub-cases by keeping at least one arrow, are absent in {\em stationary} space-times. See also \cite{S2} for similar results in space-times with all curvature scalar invariants vanishing. The compactness is essential here, as there are explicit examples of trapped surfaces even in Minkowski space-time, see Example 4.1 in \cite{S}, p.776. However, are all sort of non-compact (nearly) trapped surfaces realisable in a given stationary space-time? And if not, which particular types are feasible? Again, the classification seems relevant for these questions, as can be inferred from the following recent explicit example. In \cite{CV}, all marginally trapped surfaces ``with positive relative nullity'' were explicitly classified, up to isometries, in de Sitter, Minkowski, and anti de Sitter space-times (they are necessarily non-compact in the last two cases). In the language and notation introduced in the present paper, this means that all surfaces of type
$$
\swarrow^{\{(0)|(+,0)\}}
$$
(and the equivalent ones by reversing the future and past, and/or $\vec k$ and $\vec\ell$) have been explicitly identified in those space-times. Of course, many obvious questions and generalizations spring to mind, e.g. is there any important difference by considering the slightly more general cases of $\swarrow^{(0)|(+,0)}$ or $\swarrow^{(0)|(+,-)}$ surfaces? And what about the case $\swarrow^{(0)|(+,+)}$?

\subsection{$\varhexstar$-Surfaces}
\label{subsec:star}
$\varhexstar$-surfaces are defined as those $S$ with a causal mean curvature vector everywhere on $S$. In other words, $\vec H$ is not non-zero spacelike at any point of $S$. They are genuinely Lorentzian, as they can never exist in Riemannian cases, and they include many important cases such as all the nearly trapped surfaces and their subcases, the stationary surfaces, the timelike or null dual surfaces, et cetera. 

As a matter of fact, $\varhexstar$-surfaces can be split into three essentially different families, namely the stationary surfaces, the nearly trapped surfaces which are not stationary surfaces, and the rest which will be termed as {\em proper} $\varhexstar$-surfaces:
$$
\varhexstar-\mbox{surfaces}\left\{\begin{array}{c}
\mbox{nearly trapped surfaces (and all its sub-cases)} \\ \\
\mbox{proper $\varhexstar$-surfaces}
\end{array}
\right.
$$
Here the first case includes the traditional trapped surfaces treated in the previous subsection, as well as the stationary surfaces. 

Let us thus concentrate now on the proper $\varhexstar$-surfaces, which can be seen as ``surfaces with future-trapped {\em and} past-trapped portions'' and are characterized by having an everywhere causal mean curvature vector which realizes both causal orientations --- and therefore, $S$ contains necessarily, but {\em not} exclusively, stationary points with $\vec H =\vec 0$. This implies in particular that proper $\varhexstar$-surfaces are {\em at least ternary} surfaces. Hence, the simplest of them belong to the ternary class of surfaces and are either causal weakly half diverging (or converging), null untrapped, timelike dual or null dual. Graphically, these ternary $\varhexstar$-surfaces are
$$
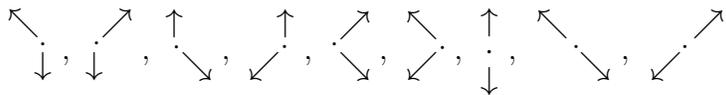
\raisebox{-3mm}{\begin{sideways}\raisebox{-1.5mm}{$\leftarrow$} \!\begin{sideways}{$\dotsearrow$}\end{sideways} \end{sideways}}\,\, ,\hspace{2mm}
\raisebox{-3mm}{\begin{sideways}\raisebox{3mm}{$\leftarrow$} $\dotsearrow$\end{sideways}}\,\, ,\hspace{2mm}
\raisebox{-3mm}{\begin{sideways}$\dotswarrow$ \raisebox{3mm}{$\rightarrow$}\end{sideways}}\,\, ,\hspace{2mm}
\raisebox{-3mm}{\begin{sideways}\begin{sideways}\begin{sideways}{$\dotsearrow$}\end{sideways}\end{sideways}
\raisebox{-1mm}{$\rightarrow$}\end{sideways}} \,\, ,\hspace{2mm} 
\raisebox{-3mm}{\begin{sideways}$\swarrow\dotsearrow$\end{sideways}}\,\, ,\hspace{2mm}
\raisebox{-3mm}{\begin{sideways}\begin{sideways}\begin{sideways}$\swarrow\dotsearrow$
\end{sideways}\end{sideways}\end{sideways}}\,\, ,\hspace{2mm}
\raisebox{-5mm}{\begin{sideways}$\leftarrow\cdot\rightarrow$\end{sideways}}\,\, ,\hspace{2mm}
\raisebox{-2mm}{\raisebox{5.5mm}{$\nwarrow$} $\dotsearrow$}\,\, ,\hspace{2mm}  \raisebox{-2mm}{$\dotswarrow$ \raisebox{5.5mm}{$\nearrow$}}.
$$

Very few things are known concerning proper $\varhexstar$-surfaces. We do not even know whether or not proper $\varhexstar$-spheres can be present in Minkowski space-time, a question which was put forward in \cite{MS} and may have relevance concerning the desirable monotonic properties of the ``Hawking mass''  \cite{H}. It should be remembered that this mass has played an important role in several approaches concerning the Penrose inequality \cite{Hay2,F,MMS}.

To understand this problem, let me recall that the Hawking mass of a closed surface $S$ is the number
$$
M_{H}(S) \equiv \sqrt{\frac{A_{S}}{16\pi}}\left(\frac{\chi(S)}{2}-\frac{1}{16\pi}\int_{S}g(\vec H,\vec H)\right)
$$
where $A_{S}$ is the total area of $S$. Given that closed $\varhexstar$-surfaces have a causal $\vec H$ everywhere, the last integral in this formula is always non-positive so that their $M_{H}$ is necessarily positive if they are topological spheres, or non-negative if they are tori. 

Imagine now that there is a spacelike 2-sphere in Minkowski space-time which is a proper $\varhexstar$-surface. This would imply that one can build a ``flow'' of spacelike surfaces joining this 2-sphere with infinity, where it is known that $M_{H}$ tends to the total ADM mass of the spacetime which in this case vanishes. It would follow that the Hawking mass {\em cannot} be monotonically non-decreasing along that flow, something which appears to be an undesirable property. Furthermore, from the results in \cite{MS} follows the impossibility of having $\varhexstar$-surfaces in Minkowski spacetime lying within any hypersurface orthogonal to a timelike Killing vector field.
Consequently, it seems logical to conjecture that there cannot be proper closed $\varhexstar$-surfaces in Minkowski spacetime, but as far as I am aware, this has not been proven yet. 

In general, the questions of existence of closed proper $\varhexstar$-surfaces ---specially in generic static and stationary space-times--- and of  their general properties according to the secondary classifications are open and may be of interest in several investigations.

\subsection{Higher-dimensional Lorentzian manifolds}
\label{subsec:higher}
Let me make some final comments about the possibility of generalizing the classification to higher dimensional space-times but keeping the co-dimension of $S$ fixed and equal to 2. If $\espaitemps$ is $n$-dimensional, again there are two null second fundamental forms and formulas (\ref{free}--\ref{det}) as well as the next to (\ref{det}) hold as they stand. Therefore, table \ref{t2} is valid in this general case and, more importantly, the {\em primary global classification applies as it is} in arbitrary dimension. Thus, the whole section \ref{sec:primary} remains valid for arbitrary dimension. This is the part of the classification based on the properties of the mean curvature vector or, equivalently, of the two null expansions.

Nevertheless, table \ref{t1} is no longer true. Of course, $\bm{K}_{\vec k}$ and $\bm{K}_{\vec \ell}$ are now $(n-2)\times (n-2)$ symmetric real matrices at any $x\in S$, so that they can also be algebraically classified easily as they are always diagonalizable with respect to the first fundamental form $\bm{\gamma}$. Letting $\{\lambda_{i}\}$ and $\{\mu_{i}\}$ ($i=1,\dots ,n-2$) be the eigenvalues of $\bm{K}_{\vec \ell}|_{x}$ and $\bm{K}_{\vec k}|_{x}$ respectively, one can order them according to 
$$
|\lambda_{1}|\geq |\lambda_{2}|\geq \dots \geq |\lambda_{n-2}|
$$
and similarly for $\mu_{i}$. Then, a symbol for each type of point on the surface would be of the type
$$
\left(\mbox{sign}(\lambda_{1}),\dots ,\mbox{sign}(\lambda_{n-2}) |\mbox{sign}(\mu_{1}),\dots ,\mbox{sign}(\mu_{n-2})\right)_{\alpha_{1},\dots,\alpha_{m}}
$$
where all signs belong to $\{+,-,0\}$ and the $\alpha_{B}$ ($B\in\{1,\dots ,m=(n-2)(n-3)/2\}$) are the angles of the $SO(n-2)$-rotation matrix relating
the corresponding orthonormal eigenbases $\{e_{i}\}$ and $\{E_{i}\}$ with the same orientation. Observe that one can use the symbol $\pm$ for any pair of eigenvalues with the same magnitude but opposite sign. There may arise many degenerate cases (corresponding to eigenvalues of multiplicity greater than one), so that the pure algebraic classification at a fixed point has increasing complexity as $n$ increases. Notice also that one can also use a set of $n-2$ invariants for each of the matrices $\bm{K}_{\vec k}$ or $\bm{K}_{\vec \ell}$ for the same purposes.

It is easily checked that the method is the same as the one that has been used here, even though the classification is much more complicated in general $n$. Thus, the secondary global classifications can follow the same procedure by simply splitting first between pure, binary, ternary, et cetera, surfaces, and then adding the corresponding type of less-specialized point, and so on, to the letters of the acronym or to the arrows in the graphical symbols.

\section*{Appendix}
In this Appendix the causal orientation of $\vec{\bm{K}}|_{x}(\vec v,\vec v)$ for $\vec v\in T_x S$ is analyzed thoroughly, for any single possible case arising from the classification in Table \ref{t2} and the subclassifications, supplemented with the possible values of $\alpha$, appearing at the six corresponding boxes of Table \ref{t3}. 

\subsection*{Expanding}
The 9 classes ---supplemented with the particular value of $\alpha$--- appearing at the left upper box of Table \ref{t3} lead to the following possibilities:
\begin{enumerate}
\item\label{1} Case $(+,+)|(+,+)_{\alpha}$: for all values of $\alpha$ and for all possible $\vec v \in T_{x}S$, $\vec{\bm{K}}|_x(\vec v,\vec v)$ is past-pointing timelike.
\item\label{2} Cases $(+,+)|(+,0)_{\alpha}$ and $(+,0)|(+,+)_{\alpha}$:  for all values of $\alpha$ and for all directions $\vec v$ {\em but one}, $\vec{\bm{K}}|_x(\vec v,\vec v)$ is past-pointing timelike. The exception is given by the eigenvector $\vec e_{2}$ (or $\vec E_{2}$) with vanishing eigenvalue, at which it is past-pointing null.
\item\label{3} Case $(+,0)|(+,0)_{\alpha}$: for all $\alpha\neq 0$ there are exactly two directions, $\vec e_{2}$ and $\vec E_{2}$, such that 
$\vec{\bm{K}}|_x(\vec e_{2},\vec e_{2})$ and $\vec{\bm{K}}|_x(\vec E_{2},\vec E_{2})$ are past-pointing null, otherwise $\vec{\bm{K}}|_x(\vec v,\vec v)$ is past-pointing timelike. 

For the congruent case $\alpha =0$, $(+,0)|(+,0)_{0}$, there is a unique direction, given by 
$\vec e_{2}= \vec E_{2}$ such that $\vec{\bm{K}}|_x(\vec e_{2},\vec e_{2})$ actually vanishes, and $\vec{\bm{K}}|_x(\vec v,\vec v)$ is past-pointing timelike for all other  
$\vec v \in T_{x}S$.
\item\label{4} Cases $(+,+)|(+,-)_{\alpha}$ and $(+,-)|(+,+)_{\alpha}$: take the case $(+,+)|(+,-)_{\alpha}$ (say). Now, for arbitrary $\alpha$, there is a continuous set of directions, given by (\ref{v}) with $\sin^2\beta > B_{c}=\lambda_{1}/(\lambda_{1}-\lambda_{2})$, such that $\vec{\bm{K}}|_x(\vec v,\vec v)$ is spacelike. At the critical values $\sin^2\beta = B_{c}$ these are past-pointing null. Otherwise, $\vec{\bm{K}}|_x(\vec v,\vec v)$ is past-pointing timelike. Observe that the values of $\beta$ such that the latter case holds are given by $|\beta|<\arcsin\sqrt{B_{c}}$, and therefore these are the dominant directions as $\arcsin\sqrt{B_{c}}>\pi/4$. 

Analogously for the other case $(+,-)|(+,+)_{\alpha}$.

\item\label{5} Cases $(+,-)|(+,0)_{\alpha}$ and $(+,0)|(+,-)_{\alpha}$: concentrating on the case $(+,0)|(+,-)_{\alpha}$, three possibilities must be distinguished according to whether 
\be
|\alpha|< \alpha_{c}\equiv \pi/2 -\arcsin \sqrt{B_{c}}\label{alphac}
\ee
or not. 
\begin{itemize}
\item If $\alpha\notin [-\alpha_{c},\alpha_{c}]$, then $\vec E_{2}$ is {\em not} one of the vectors (defined by (\ref{v}) with $\sin^2\beta > B_{c}$) such that $\bm{K}_{\vec k}|_{x}(\vec v,\vec v)<0$. Then, for those vectors, $\vec{\bm{K}}|_x(\vec v,\vec v)$ is spacelike, while it is past-pointing timelike for all other vectors {\em except} for $\vec E_{2}$ and for (\ref{v}) with $\sin^2\beta = B_{c}$, when it is past-pointing null.
\item If $\alpha\in (-\alpha_{c},\alpha_{c})$, $\vec E_{2}$ is one of the vectors such that $\bm{K}_{\vec k}|_{x}(\vec v,\vec v)<0$. For these vectors $\vec{\bm{K}}|_x(\vec v,\vec v)$ is spacelike {\em except} for $\vec E_{2}$, in which case 
$\vec{\bm{K}}|_x(\vec E_{2},\vec E_{2})$ is {\em future}-pointing null. For the remaining vectors (\ref{v}), $\vec{\bm{K}}|_x(\vec v,\vec v)$ is past-pointing timelike or null if $\sin^2\beta$ is smaller than or equal to $B_{c}$, respectively.
\item The critical cases  $(+,0)|(+,-)_{\pm\alpha_{c}}$ have the same properties as the previous one except that now $\vec E_{2}$ is parallel to (\ref{v}) with $\sin^2\beta=B_{c}$ and, actually, $\vec{\bm{K}}|_x(\vec E_{2},\vec E_{2})$ vanishes.
\end{itemize}
Analogously for the case $(+,-)|(+,0)_{\alpha}$. 

\item\label{6} Case $(+,-)|(+,-)_{\alpha}$: this is the most intricate case. There is a set of directions, given by  (\ref{v}) with $\sin^2\beta < B_{c}$, such that $\bm{K}_{\vec k}|_{x}(\vec v,\vec v)>0$; analogously, the set of directions
\be
\vec w =\cos\gamma\, \vec E_{1} +\sin\gamma\, \vec E_{2}\label{w}
\ee
with $\sin^2\gamma < \mu_{1}/(\mu_{1}-\mu_{2})\equiv C_{c}$ satisfies 
$\bm{K}_{\vec \ell}|_{x}(\vec w,\vec w)>0$. Thus, there {\em always} exists a set of common directions such that both of them are positive because $B_{c},C_{c}>1/2$ ($\arcsin \sqrt{B_{c}} + \arcsin \sqrt{C_{c}} > \pi/2$.) This set of common directions may be disconnected and for these directions $\vec{\bm{K}}|_{x}(\vec v,\vec v)$ is past-pointing timelike. There arise several cases and subcases depending on the sign of $(\arcsin \sqrt{B_{c}} - \arcsin \sqrt{C_{c}})$ and on the particular value of $\alpha$. These are:
\begin{itemize}
\item If $\arcsin \sqrt{B_{c}} < \arcsin \sqrt{C_{c}}$, one has to consider the following subcases
\begin{itemize}
\item If 
$$
\pi-\arcsin \sqrt{B_{c}} - \arcsin \sqrt{C_{c}}< |\alpha|\leq \pi/2
$$ 
then $\vec{\bm{K}}|_{x}(\vec v,\vec v)$ is spacelike for (\ref{v}) with all values of $\beta$ satisfying
$$
\pi/2\geq |\beta|> \arcsin\sqrt{B_{c}}\hspace{1cm} \mbox{or}  \hspace{7mm}
\arcsin\sqrt{C_{c}} < |\beta -\alpha| < \pi -\arcsin\sqrt{C_{c}}\, .
$$
At the boundaries $\vec{\bm{K}}|_{x}(\vec v,\vec v)$ becomes past-pointing null. 
\item If
$$
\alpha =\pm \left(\pi-\arcsin \sqrt{B_{c}} - \arcsin \sqrt{C_{c}}\right)
$$
everything happens as in the previous subcase except that 
$\vec{\bm{K}}|_{x}(\vec v,\vec v)$ vanishes for the values $\beta =\mp\arcsin\sqrt{B_{c}}$, respectively.
\item If
$$
\arcsin \sqrt{C_{c}} - \arcsin \sqrt{B_{c}}< |\alpha|
< \pi-\arcsin \sqrt{B_{c}} - \arcsin \sqrt{C_{c}}
$$
then $\vec{\bm{K}}|_{x}(\vec v,\vec v)$ is spacelike for (\ref{v}) with either
\bea
&\alpha > 0&\left\{
\begin{array}{c} \arcsin \sqrt{B_{c}}<\beta <\min\{\alpha+\arcsin\sqrt{C_{c}}, \pi /2\}\\
-\pi/2< \beta < \max\{-\pi/2,\alpha+\arcsin \sqrt{C_{c}}-\pi\} \\
-\arcsin \sqrt{B_{c}}<\beta < \alpha - \arcsin \sqrt{C_{c}}
\end{array}\right.\label{pos}\\
&\alpha < 0&\left\{
\begin{array}{c} \max\{\alpha-\arcsin \sqrt{C_{c}},-\pi /2\}<\beta < \arcsin \sqrt{B_{c}} \\
\min\{\pi/2,\pi+\alpha-\arcsin \sqrt{C_{c}}\}< \beta \leq \pi/2\\
\alpha + \arcsin \sqrt{C_{c}}<\beta <\arcsin \sqrt{B_{c}}
\end{array}\right.\label{neg}
\eea
Furthermore, now there appears a non-empty set of $\vec v\in T_{x}S$ such that $\vec{\bm{K}}|_{x}(\vec v,\vec v)$ is {\em future}-pointing timelike, given by (\ref{v}) with
\bean
\left. \begin{array}{c}
\max\{-\pi/2,\alpha+\arcsin \sqrt{C_{c}}-\pi\}<\beta < -\arcsin \sqrt{B_{c}}\\
\min\{\alpha+\arcsin\sqrt{C_{c}}, \pi /2\}<\beta <\pi/2
\end{array}\right\}(\mbox{if}\,\, \alpha >0),\\
\left. \begin{array}{c}
\arcsin \sqrt{B_{c}} <\beta < \min\{\pi/2,\pi+\alpha-\arcsin \sqrt{C_{c}}\}\\
\max\{\alpha-\arcsin \sqrt{C_{c}},-\pi /2\}< \beta<-\pi/2 
\end{array}\right\} (\mbox{if}\,\, \alpha <0)\, .
\eean
When $\alpha>0$, $\vec{\bm{K}}|_{x}(\vec v,\vec v)$ is future-pointing null at the two boundaries $\beta =\alpha+\arcsin \sqrt{C_{c}}$ (or $\beta =\alpha+\arcsin \sqrt{C_{c}}-\pi$), $\beta=-\arcsin\sqrt{B_{c}}$, while it is past-pointing null at the boundaries $\beta =\alpha - \arcsin \sqrt{C_{c}}$ (or $\beta=\pi+\alpha - \arcsin \sqrt{C_{c}}$) and $\beta = \arcsin \sqrt{B_{c}}$; and analogously for the case $\alpha <0$.
\item If
$$
\alpha = \pm \left(\arcsin \sqrt{C_{c}} - \arcsin \sqrt{B_{c}}\right)
$$
everything is as in the previous subcase except that the third possibility in (\ref{pos},\ref{neg}) is now impossible, and $\vec{\bm{K}}|_{x}(\vec v,\vec v)$ actually vanishes for the value of $\beta = -\arcsin \sqrt{B_{c}}=\alpha - \arcsin \sqrt{C_{c}}$ (if $\alpha >0$), or the value $\beta =\alpha + \arcsin \sqrt{C_{c}}=\arcsin \sqrt{B_{c}}$ (if $\alpha <0$).
\item Finally, if
$$
|\alpha|< \arcsin \sqrt{C_{c}} - \arcsin \sqrt{B_{c}}
$$
then $\vec{\bm{K}}|_{x}(\vec v,\vec v)$ is spacelike for (\ref{v}) with either
\bean
&\alpha > 0&\left\{
\begin{array}{c} \arcsin \sqrt{B_{c}}<\beta <\min\{\alpha+\arcsin\sqrt{C_{c}}, \pi /2\}\\
-\pi/2< \beta < \max\{-\pi/2,\alpha+\arcsin \sqrt{C_{c}}-\pi\} \\
\alpha - \arcsin \sqrt{C_{c}}<\beta < -\arcsin \sqrt{B_{c}}
\end{array}\right.\\
&\alpha < 0&\left\{
\begin{array}{c} \max\{\alpha-\arcsin \sqrt{C_{c}},-\pi /2\}<\beta < \arcsin \sqrt{B_{c}} \\
\min\{\pi/2,\pi+\alpha-\arcsin \sqrt{C_{c}}\}< \beta \leq \pi/2\\
\arcsin \sqrt{B_{c}}<\beta <\alpha + \arcsin \sqrt{C_{c}}
\end{array}\right.
\eean
Again there is a non-empty set of $\vec v\in T_{x}S$ such that $\vec{\bm{K}}|_{x}(\vec v,\vec v)$ is {\em future}-pointing timelike, given by (\ref{v}) with
\bean
\left. \begin{array}{c}
\max\{-\pi/2,\alpha+\arcsin \sqrt{C_{c}}-\pi\}<\beta <\alpha - \arcsin \sqrt{C_{c}}\\
\min\{\alpha+\arcsin\sqrt{C_{c}}, \pi /2\}<\beta <\pi/2
\end{array}\right\}(\mbox{if}\,\, \alpha >0),\\
\left. \begin{array}{c}
\min\{\pi/2,\pi+\alpha-\arcsin \sqrt{C_{c}}\}<\beta < \arcsin \sqrt{B_{c}} \\
\max\{\alpha-\arcsin \sqrt{C_{c}},-\pi /2\}< \beta<-\pi/2 
\end{array}\right\} (\mbox{if}\,\, \alpha <0)\, .
\eean
When $\alpha>0$, $\vec{\bm{K}}|_{x}(\vec v,\vec v)$ is future-pointing null at the two boundaries $\beta =\alpha+\arcsin \sqrt{C_{c}}$ (or $\beta =\alpha+\arcsin \sqrt{C_{c}}-\pi$), $\beta=\alpha-\arcsin \sqrt{C_{c}}$, while it is past-pointing null at the boundaries $\beta = \pm \arcsin \sqrt{B_{c}}$; and analogously for the case $\alpha <0$.
\end{itemize}
\item  Case with $\arcsin \sqrt{B_{c}} = \arcsin \sqrt{C_{c}}$. The possibilities are essentially the limit cases of the previous case. Explicitly:
\begin{itemize}
\item If 
$$
\pi-2\arcsin \sqrt{B_{c}}< |\alpha|\leq \pi/2
$$ 
then $\vec{\bm{K}}|_{x}(\vec v,\vec v)$ is spacelike for (\ref{v}) with all values of $\beta$ satisfying
$$
\pi/2\geq |\beta|> \arcsin\sqrt{B_{c}}\hspace{1cm} \mbox{or}  \hspace{7mm}
\arcsin\sqrt{B_{c}} < |\beta -\alpha| < \pi -\arcsin\sqrt{B_{c}}\, .
$$
At the boundaries $\vec{\bm{K}}|_{x}(\vec v,\vec v)$ becomes past-pointing null. 
\item If
$$
\alpha =\pm \left(\pi-2\arcsin \sqrt{B_{c}}\right)
$$
everything happens as in the previous subcase except that 
$\vec{\bm{K}}|_{x}(\vec v,\vec v)$ vanishes for the values $\beta =\mp\arcsin\sqrt{B_{c}}$, respectively.
\item If
$$
0< |\alpha| < \pi-2\arcsin \sqrt{B_{c}}
$$
then $\vec{\bm{K}}|_{x}(\vec v,\vec v)$ is spacelike for (\ref{v}) with either
\bean
&\alpha > 0&\left\{
\begin{array}{c} \arcsin \sqrt{B_{c}}<\beta <\min\{\alpha+\arcsin\sqrt{B_{c}}, \pi /2\}\\
-\pi/2< \beta < \max\{-\pi/2,\alpha+\arcsin \sqrt{B_{c}}-\pi\} \\
-\arcsin \sqrt{B_{c}}<\beta < \alpha - \arcsin \sqrt{B_{c}}
\end{array}\right.\\
&\alpha < 0&\left\{
\begin{array}{c} \max\{\alpha-\arcsin \sqrt{B_{c}},-\pi /2\}<\beta < \arcsin \sqrt{B_{c}} \\
\min\{\pi/2,\pi+\alpha-\arcsin \sqrt{B_{c}}\}< \beta \leq \pi/2\\
\alpha + \arcsin \sqrt{B_{c}}<\beta <\arcsin \sqrt{B_{c}}
\end{array}\right.
\eean
$\vec{\bm{K}}|_{x}(\vec v,\vec v)$ is {\em future}-pointing timelike for the 
$\vec v\in T_{x}S$ is (\ref{v}) with
\bean
\left. \begin{array}{c}
\max\{-\pi/2,\alpha+\arcsin \sqrt{B_{c}}-\pi\}<\beta < -\arcsin \sqrt{B_{c}}\\
\min\{\alpha+\arcsin\sqrt{B_{c}}, \pi /2\}<\beta <\pi/2
\end{array}\right\}(\mbox{if}\,\, \alpha >0),\\
\left. \begin{array}{c}
\arcsin \sqrt{B_{c}} <\beta < \min\{\pi/2,\pi+\alpha-\arcsin \sqrt{B_{c}}\}\\
\max\{\alpha-\arcsin \sqrt{B_{c}},-\pi /2\}< \beta<-\pi/2 
\end{array}\right\} (\mbox{if}\,\, \alpha <0)\, .
\eean
When $\alpha>0$, $\vec{\bm{K}}|_{x}(\vec v,\vec v)$ is future-pointing null at the two boundaries $\beta =\alpha+\arcsin \sqrt{B_{c}}$ (or $\beta =\alpha+\arcsin \sqrt{B_{c}}-\pi$), $\beta=-\arcsin\sqrt{B_{c}}$, while it is past-pointing null at the boundaries $\beta =\alpha - \arcsin \sqrt{B_{c}}$ (or $\beta=\pi+\alpha - \arcsin \sqrt{B_{c}}$) and $\beta = \arcsin \sqrt{B_{c}}$; similarly for $\alpha <0$.
\item In the congruent case $\alpha = 0$, $\vec{\bm{K}}|_{x}(\vec v,\vec v)$ is future pointing timelike for $\arcsin\sqrt{B_{c}}<|\beta|\leq \pi/2$, vanishes at $\beta=\pm\arcsin\sqrt{B_{c}}$ and is past pointing timelike for the remaining $\beta$'s.
\end{itemize}

\item The case with $\arcsin \sqrt{B_{c}} > \arcsin \sqrt{C_{c}}$ is qualitatively equivalent to the case with $\arcsin \sqrt{C_{c}}>\arcsin \sqrt{B_{c}}$ by just changing $\alpha \rightarrow -\alpha$ and $\arcsin \sqrt{B_{c}} \leftrightarrow \arcsin \sqrt{C_{c}}$ (and, if necessary, $\bm{K}_{\vec k}\leftrightarrow \bm{K}_{\vec\ell}$).
\end{itemize}
\end{enumerate}

\subsection*{Semi-Expanding}
The 6 classes at the central upper box in Table \ref{t3} lead now to the following:
\begin{enumerate}
\item Case $(+,+)|(0)$: Obviously, $\vec{\bm{K}}|_{x}(\vec v,\vec v)$ is past-pointing null for all possible $\vec v$.
\item Case $(+,+)|(\pm)_{\alpha}$: This is the limit case with $B_{c}=1/2$ of the expanding case $(+,+)|(+,-)_{\alpha}$. Observe that now there is no dominant orientation, because $\vec{\bm{K}}|_{x}(\vec v,\vec v)$ is spacelike or timelike for the $\vec v$ in (\ref{v}) with $|\beta|>\pi /4$ or $|\beta| < \pi/4$, respectively. At the boundary $|\beta|=\pi /4$, it is past-pointing null.
\item Case $(+,0)|(0)$: $\vec{\bm{K}}|_{x}(\vec v,\vec v)$ is past-pointing null for all 
$\vec v\neq \vec E_{2}$, and $\vec{\bm{K}}|_{x}(\vec E_{2},\vec E_{2})=0$.
\item Case $(+,0)|(\pm)_{\alpha}$: again, this is the limit with $B_{c}=1/2$ of the expanding case $(+,0)|(+,-)_{\alpha}$.
\item Case $(+,-)|(0)$: $\vec{\bm{K}}|_{x}(\vec v,\vec v)$ is past-pointing null, future-pointing null, or zero if $|\beta|$ is lower, greater or equal  than $\arcsin\sqrt{C_{c}}$, respectively.
\item Case $(+,-)|(\pm)_{\alpha}$: once more, this is the limit with $B_{c}=1/2$ of the expanding case $(+,-)|(+,-)_{\alpha}$.
\end{enumerate}

A similar list arises for the left central box, changing the roles of $\vec k$ and $\vec \ell$.

\subsection*{Mixed}
The 9 classes at the lower left box of Table \ref{t3} behave analogously to the 9 expanding cases by letting all the $+$'s before the bar ``$|$'' to be $-$'s, and vice versa. Then, one simply has to replace the $\vec{\bm{K}}|_{x}(\vec v,\vec v)$ which are timelike with the spacelike ones, the spacelike with timelike ones, changing the sign of the $\vec{\bm{K}}|_{x}(\vec v,\vec v)$ which are proportional to $\vec k$ but keeping those proportional to $\vec \ell$. Now, however, the dominant orientation for the types 
$(-,+)|(+,-)_{\alpha}$ with $|\alpha|\geq \pi-\arcsin\sqrt{B_{c}}-\arcsin\sqrt{C_{c}}$ can be future or past timelike, or none. The results are presented in Table \ref{t7}.

The list for the mixed cases of the right upper box in Table \ref{t3} are obtained from the previous ones by interchanging the roles of $\vec k$ and $\vec \ell$.

\subsection*{Stationary}
The 4 classes at the central box in Table \ref{t3} are quite simple to handle. Obviously the type $(0)|(0)$ is trivial and $\vec{\bm{K}}|_{x}(\vec v,\vec v)=\vec 0$ for all $\vec v$ in that case. The cases $(0)|(\pm)$ and $(\pm)|(0)$ are also very simple, as the vectors $\vec{\bm{K}}|_{x}(\vec v,\vec v)$ are proportional to $\vec \ell$ and to $\vec k$, respectively. The future or past orientation corresponds to $\vec v$ in (\ref{v}) with $|\beta|$ lower or greater than $\pi/4$, respectively. For $\beta =\pm \pi/4$, they vanish. Observe that there is no dominant orientation in these two types, as the ranges of $\beta$ for the future and past cases are equivalent. 

Finally, for the types $(\pm)|(\pm)_{\alpha}$ it is enough to consider the cases with $\alpha\geq 0$ (the cases with negative $\alpha$ are symmetric). For the congruent case, $(\pm)|(\pm)_{0}$, $\vec{\bm{K}}|_{x}(\vec v,\vec v)$ are past-pointing timelike for $|\beta|<\pi/4$, future-pointing timelike for $|\beta|>\pi/4$, and vanishing at 
$\beta =\pm \pi/4$.

For $0<\alpha \leq \pi/2$, $\vec{\bm{K}}|_{x}(\vec v,\vec v)$ are future-pointing timelike if $\max\{-\pi/2,\alpha-3\pi/4\}<\beta<-\pi/4$ and $\max\{\pi/4,\alpha+\pi/4\}<\beta< \pi/2$; past-pointing timelike if $\alpha-\pi/4<\beta<\pi/4$; spacelike if $-\pi/2<\beta<\max\{-\pi/2,\alpha-3\pi/4\}$, $-\pi/4<\beta <\alpha-\pi/4$ and $\pi/4<\beta <\min\{\pi/2,\alpha+\pi/4\}$; past-pointing null at $\beta =\alpha-\pi/4$ and $\beta=\pi/4$; and future-pointing null at $\beta=-\pi/4$ and $\beta =\alpha+\pi/4$. Observe that, for $\pi/6< \alpha\leq \pi/2$ the  spacelike orientations ($\leftarrow$ and $\rightarrow$) {\em taken together} are dominant, as there are more directions in these two classes than in any one of the others.

Finally, for the orthogonal case, $(\pm)|(\pm)_{\pi/2}$, 
$\vec{\bm{K}}|_{x}(\vec v,\vec v)$ are spacelike for all $\beta\neq \pm\pi/4$, and vanishing at $\beta =\pm \pi/4$.

\subsection*{Semi-Contracting and Contracting}
These cases are equivalent to the semi-expanding and the expanding cases, respectively, by just reversing the time orientation of $\vec{\bm{K}}|_{x}(\vec v,\vec v)$, that is to say, by interchanging all future-pointing possibilities with past-pointing ones and conversely.

\section*{Acknowledgements}
Thanks to L.J. Al\'{\i}as, A. Garc\'{\i}a-Parrado, I. Eriksson, M. Mars, and M. S\'anchez for reading the manuscript and some corrections.
Financial support under
grants FIS2004-01626 of the Spanish CICyT and 
GIU06/37 of the University of the Basque 
Country is acknowledged.

\end{document}